\documentclass[fleqn,usenatbib]{mnras}

\usepackage{newtxtext,newtxmath}
\usepackage[T1]{fontenc}
\usepackage{ae,aecompl}

\usepackage{graphicx}	
\usepackage{times}
\usepackage{amsmath}	
\usepackage{geometry}
\newcommand{\fesctd}{\mbox{$f_{\rm esc}^{\rm 3D}$}}
\newcommand{\fescod}{\mbox{$f_{\rm esc}^{\rm 1D}$}}
\newcommand{\fescrel}{\mbox{$f_{\rm esc,rel}^{\rm 1D}$}}
\newcommand{\fescreltd}{\mbox{$f_{\rm esc,rel}^{\rm 3D}$}}

\newcommand{\fesctdL}{\mbox{$\left<f_{\rm esc}^{\rm 3D}\right>_{\mathcal L}$}}

\newcommand{\fesc}{\mbox{$f_{\rm esc}$}}
\newcommand{\msun}{\mbox{$M_{\odot}$}}
\newcommand{\msunyr}{\mbox{$M_{\odot}\,{\rm yr^{-1}}$}}

\newcommand{\mvir}{\mbox{$M_{\rm vir}$}}

\newcommand{\mhalo}{\mbox{$M_{\rm halo}$}}
\newcommand{\fclump}{\mbox{$f_{\rm clump}^{\gamma}$}}

\newcommand{\tage}{\mbox{$t_{\rm age}$}}
\newcommand{\tsim}{\mbox{$t_{\rm sim}$}}
\newcommand{\tensh}{\mbox{$t_{\rm enshr}$}}

\let\oldAA\AA
\renewcommand{\AA}{\text{\normalfont\oldAA}}

\title[Origin of inefficient LyC escape in SFGs]{On the origin of low escape fractions of ionizing radiation from massive star-forming galaxies at high redshift}

\author[Yoo, Kimm, \& Rosdahl]{Taehwa Yoo$^{1}$\thanks{astro.taehwa.yoo@gmail.com}, Taysun Kimm$^{1}$\thanks{tkimm@yonsei.ac.kr, corresponding author}, and Joakim Rosdahl$^{2}$
\\
$^{1}$Department of Astronomy, Yonsei University, 50 Yonsei-ro, Seodaemun-gu, Seoul 03722, Republic of Korea\\
$^{2}$Univ Lyon, Univ Lyon1, Ens de Lyon, CNRS, Centre de Recherche Astrophysique de Lyon UMR5574, F-69230 Saint-Genis-Laval, France}
\date{Accepted XXX. Received YYY; in original form ZZZ}

\pubyear{2019}
\begin{document}
\defcitealias{marchi17}{M17}
\defcitealias{steidel18}{S18}

\label{firstpage}
\pagerange{\pageref{firstpage}--\pageref{lastpage}}
\maketitle

\begin{abstract}
 The physical origin of low escape fractions of ionizing radiation derived from  massive star-forming galaxies at $z\sim3$--$4$ is not well understood. We perform idealised disc galaxy simulations to understand how galactic properties such as metallicity and gas mass affect the escape of Lyman Continuum (LyC) photons using radiation-hydrodynamic simulations with strong stellar feedback. We find that the luminosity-weighted escape fraction from a metal-poor ($Z=0.002$) galaxy embedded in a halo of mass $M_h\simeq10^{11}\,M_\odot$ is $\left<\fesctd\right>\simeq 10\,\%$. Roughly half of the LyC photons are absorbed within scales of 100 pc, and the other half is absorbed in the ISM ($\la 2\, {\rm kpc}$). When the metallicity of the gas is increased to $Z=0.02$, the escape fraction is significantly reduced to $\left<\fesctd\right>\simeq1\%$ because young stars are enshrouded by their birth clouds for a longer time. In contrast, increasing the gas mass by a factor of 5 leads to $\left<\fesctd\right>\simeq  5\, \%$ because LyC photons are only moderately absorbed by the thicker disc. Our experiments suggest that high metallicity is likely more responsible for the low escape fractions observed in  massive star-forming galaxies, supporting the scenario in which the escape fraction is decreasing with increasing halo mass. Finally, negligible correlation is observed between the escape fraction and surface density of star formation or galactic outflow rates.
 \end{abstract}




\section{Introduction}

The emergence of the Gunn-Peterson trough \citep{gunn65} in the observed spectra of quasi-stellar objects demonstrates that the Universe became transparent to Lyman Continuum (LyC) photons a billion years after the Big Bang \citep{fan01,fan06}. The Thomson optical depth measured from the cosmic microwave background signals also shows that a significant volume of the Universe was already ionized by $z\sim8$ \citep{planck-collaboration16}. These results indicate that numerous ionizing photons were produced and escape from their host dark matter halos. However, details of the propagation of LyC radiation into the intergalactic medium (IGM) remain unclear because a direct comparison of the escape fraction with observations at the epoch of reionization is not yet feasible.

Previous studies suggest that two most likely sources of the ionizing radiation are active galactic nuclei (AGN) and star-forming galaxies \citep[e.g.,][]{madau99}, among which bright AGNs are rarely observed at $z\ga4$ \citep{haehnelt01, cowie09,fontanot14}, and are only likely to be relevant for helium reionization, which occurs at $z\sim 3$--$4$  \citep[e.g.,][]{miralda-escude00,kriss01,furlanetto08,shull10,syphers14,worseck16}. \citet{giallongo15, giallongo19} argue that the number density of faint AGNs with $-22.5<M_{\rm UV}<-18.5$ is significantly higher than what was previously estimated, in which case AGNs alone could possibly explain the reionization history of the Universe \citep{madau15}. However, the ionizing emissivity estimated from recent observational surveys at $z\sim 6$ \citep{daloisio17,parsa18} is decreased by an order of magnitude compared to previous findings, in conflict with the AGN-driven reionization picture.

In contrast, star-forming galaxies can fully ionize the Universe at $z\sim6$, provided that the three-dimensional escape fraction of LyC photons is high ($\fesctd\sim10$--$20\%$) \citep[e.g.,][]{robertson13}. Such a large $\fesctd$ was also required to match the high electron optical depths ($\tau_e=0.084$) derived from observations from the Nine-year Wilkinson Microwave Anisotropy Probe \citep[e.g.][]{hinshaw13}, although the latest Planck results seem to favour a much lower $\tau_e$ of $0.056\pm0.007$ \citep{planck-collaboration18}. Indeed, results of recent cosmological radiation-hydrodynamic simulations such as SPHINX \citep{rosdahl18} have shown that galaxies with a moderate \fesctd\ of $\sim 7\%$ can fully ionize the simulated universe by $z\sim 7$ without any contribution from AGNs. 
  
 Constraining the escape of LyC photons from observations of high-z galaxies ($z\ga4$) would be extremely useful to understand reionization. \citet{kakiichi18} inferred the escape fraction of $\fesctd\sim8\%$ from $z\sim6$ Lyman-break galaxies by using the spatial correlation between the position of galaxies and Ly$\alpha$ transmission peaks in the quasar spectrum, but the direct detection of LyC flux from galaxies at $z\ga4$ is still challenging for several reasons.
First, because the density of neutral hydrogen in the IGM increases with redshift \citep[e.g.,][]{inoue08}, it is difficult to directly detect LyC flux from galaxies during the epoch of reionization. Second, dwarf galaxies that are likely to have ionized the Universe at $z>6$ are too faint to observe with the current telescope facilities. For example, \citet{kimm17} claim that galaxies embedded in  dark matter halos of mass $\mvir\la 10^{9-10}\,\msun$ are crucial to reproduce a Thomson optical depth of $\tau_e\approx0.05$--$0.06$. These objects are likely to be fainter than the UV magnitude of $M_{\rm 1500} \sim -16$ \citep{kimm14,xu16,ocvirk18}, which is close to the detection limit of large observational campaigns \citep[e.g.][]{bouwens10}. For these reasons, a sample of bright star-forming galaxies, such as Lyman-break galaxies (LBGs) or Lyman Alpha Emitters (LAEs), is often used instead to study the escape of LyC photons at inter-mediate redshifts ($z\sim 3$--$4$), although the selection method may be biased towards LyC weak galaxies \citep{cooke14}. Third, even if bona fide LyC leakers are identified, information regarding the intrinsic spectral energy distributions (SEDs) and the attenuation of LyC flux along the line-of sight is still required to convert the observed flux density ratio between ionizing and non-ionizing UV radiation, such as $F_{\rm 900}/F_{\rm 1500}$ (where the subscript indicates the average wavelength in Angstrom) to an one-dimensional absolute escape fraction (\fescod).
Note that the intrinsic flux ratios estimated using stellar population synthesis models have a broad range of $0.15\la F_{\rm 900}/F_{\rm 1500} \la 0.66$ for star-forming galaxies at $z\sim3$ \citep[e.g.,][]{inoue05,guaita16}, depending on their age and metallicity, where $F$ is in units of ${\rm erg \,s^{-1}\,Hz^{-1}}$. In addition, the properties of dust at high redshifts are not well constrained, and the wavelength-dependence of the absorption in UV wavelengths is still uncertain. Last, simulations suggest that \fescod\ measured from different orientations may significantly vary  \citep[e.g.,][]{wise09,kim13b,kimm14,paardekooper15}, especially when \fesctd\ is low \citep{cen15}.

The escape of LyC photons from high-redshift galaxies can nevertheless be measured using narrow-band or inter-mediate-band photometry  \citep{vanzella10,mostardi13,siana15,mostardi15,grazian16,grazian17,fletcher19} or spectroscopy \citep{steidel01,shapley06,marchi17,marchi18,vanzella18, steidel18}. 
However, because LyC leakers above the three $\sigma$ detection limit are rare ($\la1$--$10\%$) in the $z\sim3$--$4$ surveys, stacking analysis of the observed spectra or images has been used to measure the average or upper limit of the relative escape fraction (\fescrel), which is $\fescod$ relative to the escape fraction in non-ionizing radiation (i.e. $\lambda\approx 1500\, $\AA) after IGM correction \citep{steidel01}. For example, \citet{vanzella10} found that the 1$\sigma$ upper limit of $\fescrel$ is $<5$--$6\%$ for 102 LBGs at $3.4\la z\la4.5$. Similarly, \citet{grazian16} sampled 37 of VIMOS Ultra Deep Survey (VUDS) galaxies at $3.27\la z\la3.40$ and obtained the 1$\sigma$ upper limit of $\fescrel<2\%$ in the stacked image. Based on spectroscopic data of 33 VUDS galaxies, \citet{marchi17} also reported  $\fescrel=8-9\%$ for galaxies with $M_{\rm UV}\sim-20$ at $z\sim4$. 
Given that \fesc\ of the non-ionizing UV radiation is typically assumed to be $\sim0.2$--$0.3$ \citep[e.g.,][]{siana07}, the empirical values of \fescrel\ from the LBG and LAE samples likely indicate that their \fesctd\ is significantly lower than what is required to explain the reionization history of the Universe (i.e. $\sim10$--$20\%$). The discrepancy may be attributed to the evolution of galactic properties from $z\ga 6$ to $z\sim3$, but it may also be possible that the low \fescrel\ is related to the mass dependence, because the observed samples are biased towards bright galaxies.

Understanding the propagation of LyC photons in galaxies during the epoch of reionization has also been  attempted theoretically. \citet{gnedin08} and \citet{wise09} suggested that the LyC escape fraction is well correlated with halo mass, with higher \fesctd for more massive halos.  \citet{gnedin08} attributed this result to the fact that ionizing radiation directly escapes from the star particles located in the extended stellar disc, whereas LyC photons efficiently escaped in \citet{wise14} because star formation is more bursty in more massive halos. In contrast, \cite{razoumov10} argued that the escape fraction decreases from 80--100\% to 10\% as halo mass increases from $10^{8} M_\odot$ to $10^{11} M_\odot$. Similarly, \cite{yajima11} found that $\fesctd$ varies from $\simeq 50\%$ to $5\%$ in the halo mass range of $10^{9}$--$10^{11}\,M_\odot$. Recent studies based on radiation-hydrodynamic (radiation-hydrodynamics) simulations focusing on lower halo masses ($M_h\la 10^{10}\,\msun$) \citep[e.g.][]{wise14,kimm14,xu16,trebitsch17,kimm17} or hydrodynamic simulations with post-processing \citep[e.g.][]{paardekooper15} or semi-analytic approaches based on   observational constraints \citep[][]{finkelstein19} also reached the same conclusion, suggesting that the low escape fraction detected in bright LBGs may reflect the dependence on halo mass.

However, the physical origin of the low escape fractions is not clearly understood probably because simulating a massive system of halo mass $10^{11-12}\,M_\odot$, which is the typical host halo mass of LBGs and LAEs \citep[e.g.,][]{adelberger05,gawiser07}, with high-resolution and well-calibrated feedback models is computationally expensive and non-trivial. In this study, we attempt to unravel which physical processes or properties cause the discrepancy in escape fractions between LBGs/LAEs and the value required for reionization by performing various controlled idealised simulations. In particular, we will show how the interaction between radiation and small-scale gas clumps affects the escape of LyC photons in various environments. In Section 2, we describe the initial conditions and input physics of our radiation-hydrodynamics simulations. Section 3 presents our main results on the dependence of the escape fraction on metallicity and gas fraction. In Section 4, we compare our results with observations and discuss the connection with star formation rate density and outflow rates, including the caveat of our simulations. We summarise our findings in Section 5.

\section{Simulations}

To study the escape of LyC photons in disc galaxies, we use the \textsc{Ramses-rt} adaptive mesh refinement radiation-hydrodynamic code  \citep{teyssier02, rosdahl13, rosdahl15RT}. The Euler equations of hydrodynamics are solved with the HLLC scheme \citep{toro94} adopting a Courant number of 0.7. The radiative transfer equations are solved with a first-order moment method, the M1 closure relation for the Eddington tensor, and the GLF intercell flux function \citep{rosdahl13}. The speed of light is reduced to  1\% of the true speed of light to maintain a low computational cost while reasonably capturing the propagation of the ionization front in the dense ISM \citep[e.g.,][]{rosdahl13}. 
 
 For non-equilibrium photo-chemistry, we compute the ionization and dissociation fractions of seven species -- HI, HII, HeI, HeII, HeIII, H$_{2}$, and $e^{-}$ as described in \citet{rosdahl13} and  \citet{katz17}. 
Radiative cooling due to the primordial atomic species and molecular hydrogen is self-consistently calculated based on the non-equilibrium chemistry \citep{rosdahl13,katz17}. We also include atomic metal cooling, down to $\sim 10^4\,{\rm K}$, by adopting the cooling curves obtained from the Cloudy code \citep[\texttt{cc07} model  in \textsc{Ramses-rt},][]{ferland98}  with the UV background at $z=0$ \citep{haardt12} and fine-structure line cooling by \citet{rosen95}, down to $\sim 1 \,{\rm K}$.

\subsection{Star Formation}

  We use the thermo-turbulent scheme \citep[][Devriendt et al. {\sl in prep.}]{kimm17,kimm18} to model the formation of star particles based on a Schmidt law \citep{schmidt59}, 
  \begin{equation}
      d \rho_*/dt=\epsilon_{\rm ff} \rho_{\rm gas} / t_{\rm ff},
  \end{equation}
  where $\rho_*$ is the stellar mass density, $\rho_{\rm gas}$ is the gas density, and $\epsilon_{\rm ff}$ is the star formation efficiency per free-fall time ($t_{\rm ff}\equiv\sqrt{3\pi/32G \rho_{\rm gas}}$,  where $G$ is the gravitational constant ). The basic idea of the model is that $\epsilon_{\rm ff}$ is determined by the local thermo-turbulent conditions such that gravitationally well bound regions preferentially form stars, as suggested by the small-scale simulations of star formation \citep[e.g.,][]{padoan11,federrath12}.

  Specifically, assuming a log-normal distribution  ($p$) of gas density, the star formation rate per free-fall time may be expressed as the sum of gas mass whose density is greater than the critical density divided by free-fall time, which can be given as
  \begin{equation}
  \label{eq:sfr_ff}
      \epsilon_{\rm ff}=\frac{\epsilon_{\rm acc} }{\phi_t}\int_{s_{\rm crit}}^\infty \frac{\rho}{\rho_0}\frac{t_{\rm ff}(\rho_0)}{t_{\rm ff}(\rho)} \, p(s) \, ds,
  \end{equation}
  where $s\equiv \ln{(\rho/\rho_0)}$. Here, $\rho_0$ is the average density of the star-forming cloud, $\epsilon_{\rm acc}<1$ accounts for pre-stellar feedback processes \citep[e.g.,][]{matzner00}, and $\phi_t$ is a parameter in the order of unity which encapsulates the uncertainty in the timescale factor ($t_{\rm ff}(\rho_0)/t_{\rm ff}(\rho)$) \citep{federrath12}. 
   The critical density ($s_{\rm crit}$) can be computed by defining the boundary of a collapsing cloud with supersonic turbulence using the shock jump conditions and is given as
   \begin{equation}
      s_{\rm crit}=\ln{[0.067\theta^{-2}\alpha_{\rm vir}\mathcal{M}^2]},
  \end{equation}
  where $\theta$ is a numerical factor, $\alpha_{\text{vir}}$ is the virial parameter, and $\mathcal{M}$ is the Mach number.
  \cite{hennebelle11} and \cite{federrath12} show that for molecular gas with multi-freefall timescales, the timescale factor is no longer a constant, and Eq.~\ref{eq:sfr_ff} can be expressed as 
  \begin{equation}
  \label{eq:str_ff2}
      \epsilon_{\rm ff}=\frac{\epsilon_{\rm acc}}{2\phi_t}\exp{\left(\frac{3}{8}\sigma_s^2\right)} \left[1+{\rm erf} \left(\frac{\sigma^2_s-s_{\text{crit}}}{\sqrt{2\sigma_s^2}}\right)\right]
  \end{equation}
  where $\sigma_s^2$ is the variance in the logarithmic gas density contrast. Following \citet{federrath12}, we adopt $\epsilon_{\rm acc}=0.5$, $\theta=0.33$, and $1/\phi_t=0.57$.

 Once $\epsilon_{\rm ff}$ is determined, we evaluate the mass number ($N \equiv m_* / m_{\rm *,\, min}$), which represents the mass of a newly formed star in units of the minimum mass of a star particle based on the Poisson distribution \citep{rasera06}, with the mean of  
  \begin{equation}
      \Bar{N}=\epsilon_{\rm ff} \frac{\Delta t} {t_{\rm ff}}\frac{m_{\rm cell}}{m_{\rm *,min}},
  \end{equation}
  where $m_{\rm cell}$ is the gas mass in the cell. The minimum mass of a star particle ($m_{\rm *,min}$) is defined as 
  \begin{equation}
      m_{\rm *,min}=\frac{M_{\rm SN}n_{\rm SN}}{\eta_{\rm SN}},
  \end{equation}
  where $\eta_{\rm SNII}$ and $M_{\rm SNII}$ are the mass fraction and average progenitor mass of Type II supernova (SN), respectively, and $n_{\rm SN}$ is the minimum number of SN explosions per star particle. Note that this is necessary to model discrete, multiple SN explosions per star particle. We use $n_{\rm SN}=10$ for fiducial runs, so that each star particle has $m_{\rm *,min}=910\,\msun$ when the Kroupa initial mass function \citep[IMF, ][]{kroupa01} is assumed. 
  
\subsection{Stellar feedback}

We include five different forms of stellar feedback -- photo-ionization heating, direct radiation pressure \citep{rosdahl13}, non-thermal pressure of multi-scattering infrared photons \citep{rosdahl15RT}, and Type II SN explosions \citep{kimm14,kimm15}. To maximise the impact of feedback in low-metallicity environments, we also include the sub-grid model of radiation pressure due to multi-scattering Lyman alpha photons \citep{kimm18}. For the SEDs that each star particle emits, we use the binary population and spectral synthesis models by \citet[][BPASS v2.0]{stanway16},  which is shown to better reproduce the early reionization of the Universe than models with single stellar evolution \citep[][see \citealt{topping15} for a different choice for model SEDs]{rosdahl18,ma16,gotberg19}. We do not assume any subgrid model for the escape of LyC photons at the resolution scale, and directly compute the absorption within each cell based on the local thermodynamic properties.

\begin{table}
   \caption{Properties of the eight photon groups used in this study. From left to right, each column indicates the name, minimum and maximum energy range of photon, and the main function.}
   \centering
   \begin{tabular}{lccl}
   \hline
  Photon & $\epsilon_0$ & $\epsilon_1$  & Main function \\
  group           &      [eV]          &   [eV]   & \\
     \hline
   IR & 0.1 & 1.0  & Radiation pressure\\
   Optical & 1.0 & 5.6  &Radiation pressure\\
   FUV & 5.6 & 11.2  & Photo-electric heating\\
   LW & 11.2 & 13.6 & $\rm H_2$ Photo-dissociation\\
   EUV$_{\rm HI,1}$ & 13.6 & 15.2  & HI ionization\\
   EUV$_{\rm HI,2}$ & 15.2 & 24.59  & HI and $\rm H_2$ ionization\\
   EUV$_{\rm HeI}$ & 24.59 & 54.42  & HeI ionization\\
   EUV$_{\rm HeII}$ & 54.42 & $\infty$  & HeII ionization\\
        \hline
   \end{tabular}
   \label{tab:photongroups}
\end{table}

\subsubsection{Radiation Feedback}
Radiation feedback plays an important role in heating up and lowering the density of gas at which supernovae (SNe) explode. We use eight photon groups to model the photo-ionization of hydrogen and helium \citep{rosdahl13}, photo-dissociation of molecular hydrogen \citep{katz17}, photo-electric heating on dust \citep{kimm17}, and non-thermal pressure due to multi-scattering infrared radiation \citep{rosdahl15RT}, as summarized in Table~\ref{tab:photongroups}. Interested readers are referred to \citet{kimm18} for details, and we describe  the most important processes below for the sake of completeness.

Photo-ionization heating and direct radiation pressure are modelled by adding momentum and energy injection terms into the Euler equations \citep{rosdahl13}. 
 Dust opacity is assumed to be $5\,{\rm cm^2\,g^{-1}} \, \left( Z/Z_{\odot} \right)$ for the IR photon group and $10^3\,{\rm cm^2\,g^{-1}}\, \left(Z/Z_{\odot}\right)$ for other bands, where $Z_\odot=0.02$ is the solar metallicity. The UV and optical fluxes that are absorbed by dust or atomic species are re-radiated as IR radiation and are used to calculate the non-thermal radiation pressure due to IR photons. Note that the IR photons can freely stream and diffuse out of the source if the optical depth of dust is low \citep{rosdahl15RT}.

\begin{table*}
    \centering
    \caption{Initial conditions and set-up of performed simulations. From left to right, the columns show the minimum cell size ($\Delta x_{\rm min}$), virial radius ($R_{\rm vir}$), virial mass of a simulated halo ($M_{\rm halo}$), gas mass in the disc ($M_{\rm gas}$),  stellar mass in the disc ($M_{\rm disc,star}$) and bulge ($M_{\rm bulge,star}$),  gas metallicity ($Z_{\rm gas}$), disc gas fraction ($f_{\rm gas}\equiv M_{\rm gas} / [M_{\rm gas}+M_{\rm disc,star}]$), and the number of SN explosions per $10^3\,\msun$ stellar mass. All simulations include five different types of stellar feedback, i.e., photo-ionization heating, direct radiation pressure, non-thermal radiation pressure due to infrared and Lyman alpha photons, photo-electic heating on dust, and Type II SN explosions.}
    \begin{tabular}{lccccccccc}
    \hline
    Name  & $\Delta x_{\rm min}$  & $R_{\rm vir}$  & $M_{\rm halo}$  &   $M_{\rm gas}$ &  $M_{\rm disc,star}$ &  $M_{\rm bulge,star}$ & $Z_{\rm gas}$ & $f_{\rm  gas}$ & $N_{\rm SNII}$ \\
        &  ${\rm [pc]}$ &  ${\rm [kpc]}$ & $[M_\odot]$ &  $[M_\odot]$ & $[M_\odot]$ &  $[M_\odot]$ & & & \\
    \hline
    \texttt{G9\_Zlow} & 9.2 & 89 & $10^{11}$ &   $1.75\times10^{9}$ &$1.75\times10^{9}$ & $3.5\times10^{8}$ & 0.002 & 0.5 & 11  \\
    \texttt{G9\_Zhigh} & 9.2 & 89 & $10^{11}$ &  $1.75\times10^9$ & $1.75\times10^{9}$ & $3.5\times10^{8}$ &0.02 & 0.5 &  11   \\
        \hline
     \texttt{G9\_Zlow\_gas5} & 9.2 & 89 & $10^{11}$ &  $8.75\times10^{9}$ & $1.75\times10^{9}$ & $3.5\times10^{8}$ & 0.002 & 0.83 & 11 \\    \texttt{G9\_Zhigh\_SN5} & 9.2 & 89 & $10^{11}$ & $1.75\times10^9$ &$1.75\times10^{9}$ & $3.5\times10^{8}$ &0.02 & 0.5 & 50\\
    \texttt{G9\_Zmid\_SN5} & 9.2 & 89 & $10^{11}$ &  $1.75\times10^9$ & $1.75\times10^{9}$ & $3.5\times10^{8}$ & 0.006 & 0.5 & 50  \\
     \texttt{G9\_Zlow\_HR} & 4.6 & 89 & $10^{11}$ &  $1.75\times10^{9}$ & $1.75\times10^9$ &$3.5\times10^{8}$ & 0.002 & 0.5 & 11  \\
    \hline
    \end{tabular}

    \label{tab:setting}
\end{table*}

Resonantly scattering line emission can also contribute to the non-thermal pressure \citep[e.g.,][]{oh02,dijkstra08,aaronsmith17}. \cite{kimm18} developed a subgrid model for momentum transfer due to Lyman alpha scattering by using the Monte Carlo radiative transfer code, \textsc{rascas} \citep{rascas}. This is again performed by adding momentum to the Euler equation, based on the local multiplication factor ($M_{\rm F}$), which is defined as
\begin{equation}
    F_{Ly\alpha}=M_F\frac{L_{Ly\alpha}}{c}.
\end{equation}
 where  $F_{Ly\alpha}$ is the force, $L_{Ly\alpha}$ is the Ly$\alpha$ luminosity  (in units of ${\rm erg s^{-1}}$) originating from recombination  in each cell.  Because Ly$\alpha$ radiation interacts only with neutral hydrogen or dust, ionized hydrogen will be transparent to Ly$\alpha$ photons and no pressure will be exerted (i.e. $M_{\rm F}=0$). We adopt the  dust-to-metal ratio from \cite{remyruyer14}, and thus, the maximum values of $M_F$ for the $Z=0.002$, $Z=0.006$, and $Z=0.02$ cases are $121$, $63$, and $48$, respectively \citep[c.f.][]{smith19}.

\subsubsection{Type II supernova}

 To avoid artificial radiative losses in SN remnants due to finite resolution \citep[e.g.,][]{kim15,martizzi15}, we use the mechanical SN feedback scheme illustrated in \cite{kimm14,kimm15,kimm17}. The scheme is designed to ensure the transfer of the correct radial momentum to the surroundings by differentiating the energy-conserving and momentum-conserving phase of the Sedov-Taylor blast wave \citep{sedov59,taylor50}. In practice, the momentum of SN blast wave is calculated based on \citet{thornton98} using a mass loading factor $\chi \equiv M_{\rm swept}/M_{\rm ej} $, as 
\begin{equation}
    p_{\rm SN} = 
    \left\{
    \begin{array}{ll}
      \sqrt{2M_{\rm swept} \, f_e \, E_{\rm 51}} & \chi < \chi_{\rm tr} \\
      p_0 \,  E^{16/17}_{51} \, n_{\rm 1}^{-2/17} \,  Z_{\rm sol}^{-0.14}  &   \chi \ge \chi_{\rm tr}\\
    \end{array} 
    \right.,
\end{equation}
where $M_{\rm swept}$ is the swept-up mass, $M_{\rm ej}$ is the mass of SN ejecta, $E_{\rm 51}$ is the explosion energy of SNe in units of $10^{51}\,{\rm erg}$, $n_{\rm 1}$ is the hydrogen number density in units of ${\rm cm^{-3}}$ , $Z_{\rm sol}={\rm max}(Z/Z_{\odot},0.01)$ is the ambient gas metallicity normalised to the solar value ($Z_\odot=0.02$), and $f_e=1-\frac{(\chi-1)}{3\chi_{\rm tr}-3}$ is a  parameter that smoothly connects the two regimes. 
Here, the transition is determined as $\chi_{\rm tr}\simeq46.22 \, E_{51}^{-2/17} \, n^{-4/17}_{\rm 1} \, Z_{\rm sol}^{-0.28}$.
We use $p_0=2.5\times10^5 \, {\rm km \, s^{-1}}$, the terminal radial momentum of SN exploding at $n_{\rm H}=1\,{\rm cm^{-3}}$, which is appropriate for our radiative cooling rates.

  \begin{figure*}
    \centering
    \includegraphics[width=15cm]{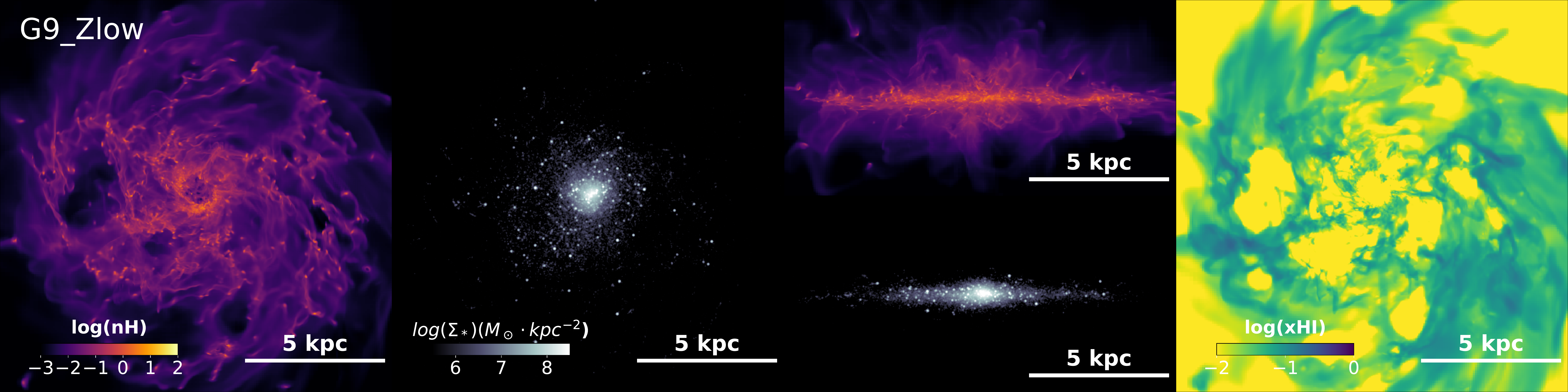}
    \includegraphics[width=15cm]{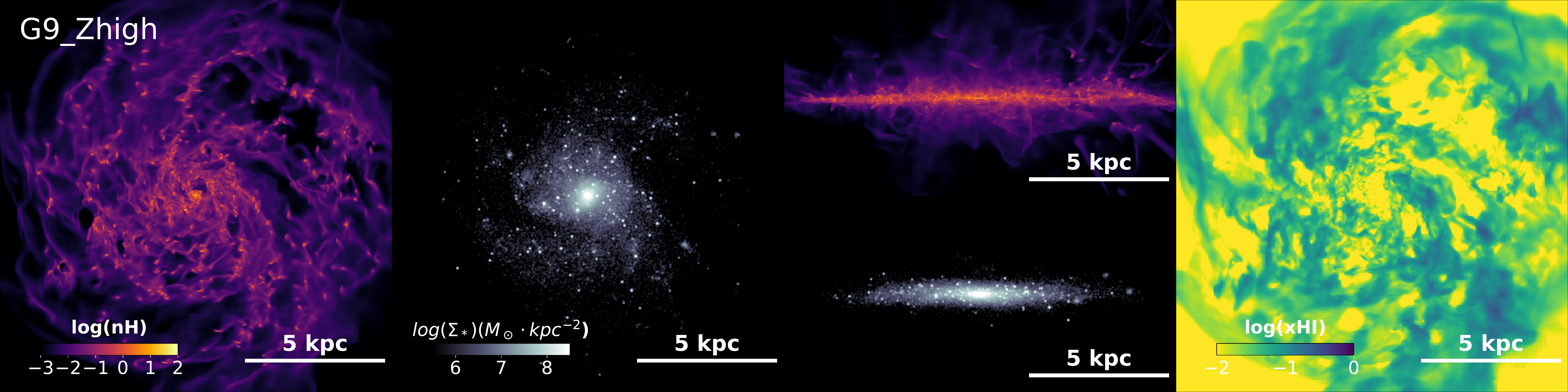}
    \includegraphics[width=15cm]{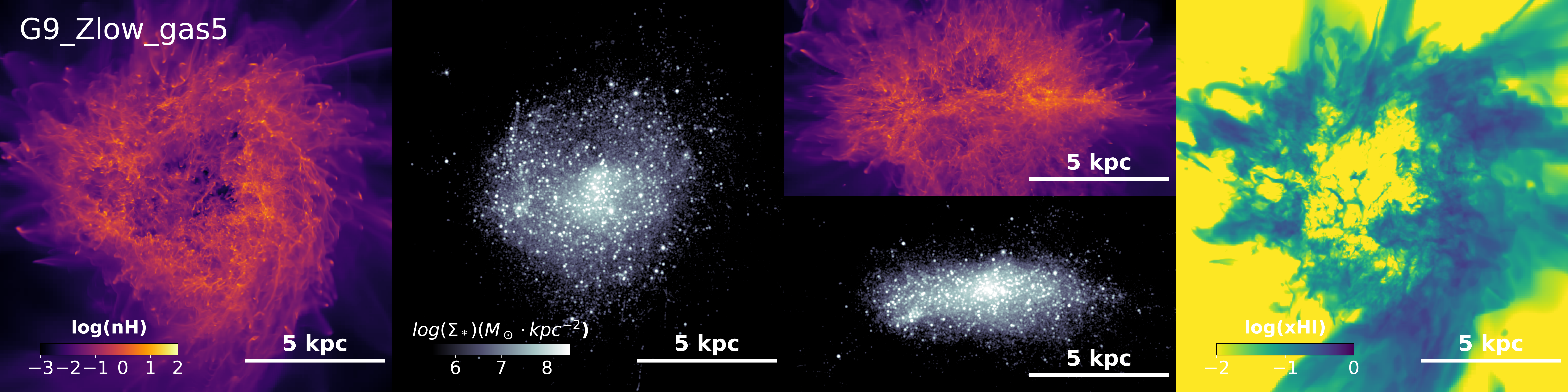}
    \includegraphics[width=15cm]{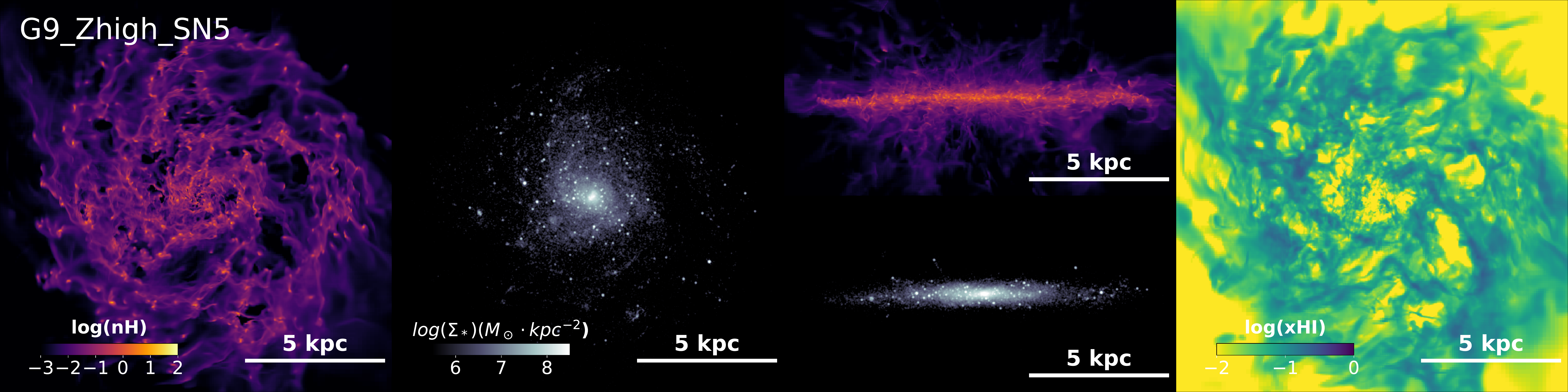}
    \includegraphics[width=15cm]{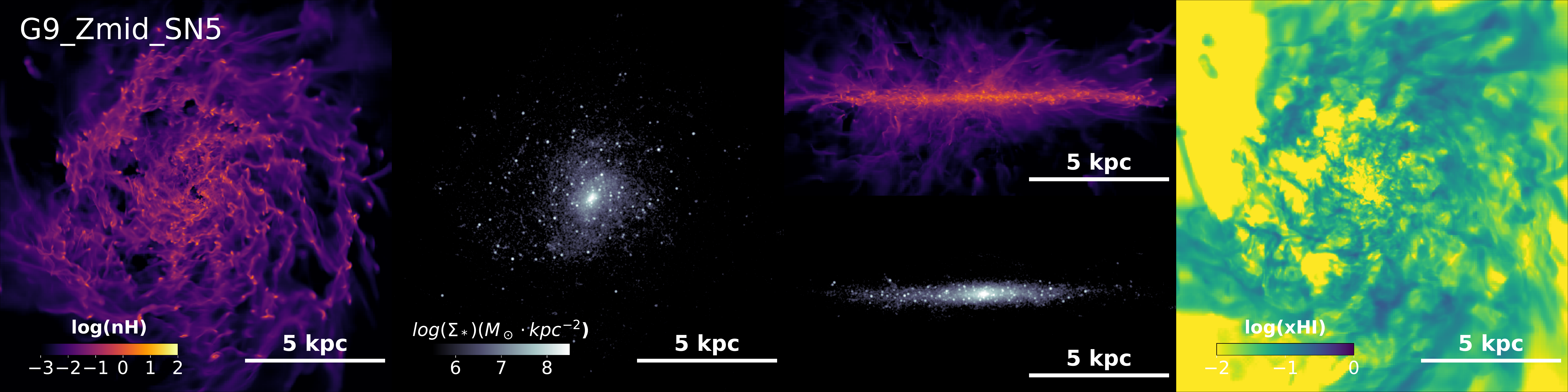}
    \includegraphics[width=15cm]{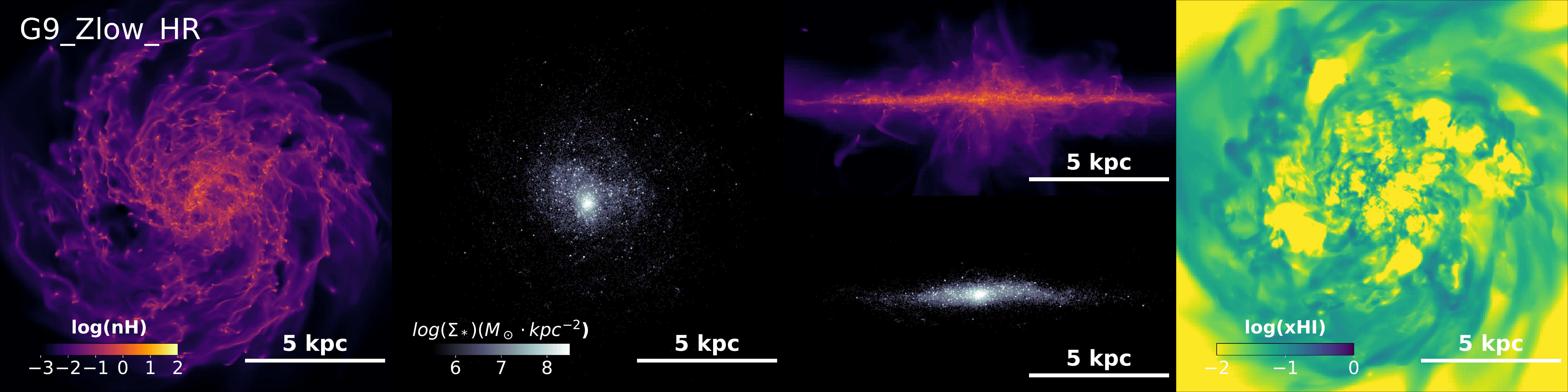}
    
    \caption{Images of the simulated disc galaxies at their final snapshot. The images measure $7\,{\rm kpc}$ on a side. The first two columns show the face-on projected distributions of hydrogen number density and stellar surface density, the third column displays the edge-on distributions of gas (upper) and stars (lower panel), and the fourth column shows projections of the mass-weighted average fraction of neutral hydrogen (depth of $200\,{\rm pc}$).}
    \label{fig:fig1}
\end{figure*}

Each star particle hosts at least 11 SN explosions between 4 and 40 Myr after its birth, which are modelled as multiple discrete events. The lifetime of SN progenitors of mass $8 \le M \le 100 \,\msun$ is randomly sampled using the inverse method, as described in \cite{kimm15}. The total mass fraction returned to the ambient medium via SN explosions is $\eta_{\rm SN}=0.21$ for the fiducial runs, and we use the mean progenitor mass of $M_{\rm SN}=19.1\,\msun$, which is appropriate for the Kroupa IMF. We also perform runs with extreme SN feedback in which the number of SN per unit stellar mass is boosted by $4.5$. This is motivated by the fact that cosmological simulations require 4--5 times stronger SN feedback than the fiducial cases to reproduce the stellar-to-halo mass relation, the UV luminosity functions at $z=6$ \citep{rosdahl18}, and/or the star formation history of Milky Way-like galaxies \citep{li18}. For these extreme feedback models, denoted with  \texttt{SN5}, we use $\eta_{\rm SN}=0.2$ and $M_{\rm SN}=4\,\msun$.

\subsection{Initial conditions and simulation set-up}

We employ the initial conditions of the G9 simulation from  \cite{rosdahl15shine} to study the escape of LyC photons from disc galaxies  of stellar masses $\sim10^9M_\odot$ embedded in a dark matter halo of mass $\mhalo =10^{11} \, M_\odot$, which is the typical halo mass of LAEs \citep[e.g.,][]{gawiser07} or a lower limit of the host mass of LBGs \citep[e.g.,][]{adelberger05}. The simulated box width is set to $300 \, {\rm kpc}$ on a side to cover the entire virial radius of $89\,{\rm kpc}$. At the center of the box, we place the stellar disc of mass $1.75\times10^9\,\msun$, stellar bulge of mass $3.5\times10^8\,\msun$, and gaseous disc of mass $M_{\rm gas}=1.75\times10^9\,\msun$ (\texttt{G9\_Zlow}). These pre-existing stars do not produce ionizing or non-ionizing photons, but only interact via gravity.  The corresponding disc gas fraction, defined as the disc gas mass divided by the total disc mass ($f_{\rm gas}\equiv M_{\rm gas}/[M_{\rm gas}+M_{\rm disc, star}])$, is 0.5, which is the typical gas fraction derived from LBGs at $z\sim3$ \citep{schinnerer16}. We also run a simulation with five times larger $M_{\rm gas}$ ($f_{\rm gas}=0.83$) than the fiducial case to see the effects of gas mass  (\texttt{G9\_Zlow\_gas5}). A factor of five is chosen to bracket the maximum gas fractions inferred from the observations of LBGs \citep{schinnerer16}.

 High-$z$ star-forming galaxies of stellar mass $M_*\simeq10^{9-11}\,M_\odot$ have a broad range of metallicities with $Z=0.1$--$1Z_\odot$ \citep{mannucci09,onodera16}. Motivated by this and also to examine the effects of metallicity, we set the gas metallicity of the fiducial and metal-rich runs (\texttt{G9\_Zhigh\_XXX}) to $Z=0.002$ and $Z=0.02$, respectively. Because the typical metallicity of LBGs at $z\sim3$--$4$ is $\sim 0.3$--$0.4\,Z_\odot$ \citep[e.g.][]{mannucci09}, we run an additional case with $Z=0.006$ (\texttt{G9\_Zmid\_XXX}) to check if the simulated galaxy with typical metallicity produces escape fractions consistent with the observations. Note that we set the stellar yield to zero in all runs to avoid any possible confusion due to metal enrichment during the evolution of our simulations.

The simulated volume is covered with $128^3$ coarse cells, and these are further refined if the gas mass in a cell exceeds $1000\,\msun$ or if the local Jeans length is resolved by fewer than four cell widths \citep{truelove97} until it reaches the maximum spatial resolution of $\Delta x_{\rm min}=9.2\,{\rm pc}$ (refinement level 15). The fiducial galaxy is typically resolved by $\sim$ 18 million leaf cells, of which $\sim $ 5 million cells are at the maximum refinement level. For comparison, the gas-rich run has $\sim 80$ million leaf cells out of which $\sim20$ million cells are maximally refined. We also run a simulation with one more level of refinement, i.e. $\Delta x_{\rm min}=4.6\, {\rm pc}$, to check for resolution convergence (\texttt{G9\_Zlow\_HR}). Five simulations with the fiducial gas fraction are run until $\tsim\approx 490\,{\rm Myr}$, whereas the gas-rich run is stopped at $\tsim\approx 290 \,{\rm Myr}$ due to limited computational resources. We output the snapshots with 1 Myr intervals  at $\tsim>150\,{\rm Myr}$ for accurate measurements of the escape fractions. The simulation set-up is summarised in Table~\ref{tab:setting}, and the corresponding images of the simulated disc galaxies at their final snapshot are shown in Fig.~\ref{fig:fig1}.

\subsection{Measurement of the escape fraction}

The escape fraction of LyC photons is defined by the ratio of the total number of LyC photons produced inside a galaxy and the number of LyC photons escaping to the virial radius. 
In principle, the escape fraction can be measured directly by comparing the ionizing flux generated by star particles and the flux reaching the virial sphere \citep[e.g.,][]{kimm14}. However, as we are interested in measuring the scale at which the majority of LyC photons are absorbed from each source,  we post-process the snapshots with a simple ray-tracing method to make the best use of our simulations as follows\footnote{\citet[][see their Figure 7]{trebitsch17} demonstrated that the escape fractions measured from the post-processing with the ray-tracing method are in good agreement with those obtained from the method based on fluxes when a speed of light of $10^{-2}\, c$ is used.}.

To measure the escape fraction, we calculate the optical depth for each star particle along 768 directions using the {\sc HEALPix} tessellation algorithm \citep{gorski05}. We first assign an SED to each ray, based on the age and metallicity of stellar population \citep{stanway16}, and trace the ray until it reaches the virial radius of the host dark matter halo ($R_{\rm vir}=89\,{\rm kpc}$). 

Attenuation due to hydrogen and helium is computed as a function of wavelength by adopting the absorption cross-sections specified in \citet{osterbrock06} for $\rm HI$ and $\rm He II$  and the absorption cross-sections from \citet{yan98,yan01} for $\rm H_{\rm 2}$ and $\rm He  I$, respectively. We also consider the dust opacity based on the Small Magellanic Cloud-type dust from \cite{weingartner01}. The dust-to-metal ratio is assumed to be 0.4 for the neutral ISM, whereas a lower value is adopted in hot ionized regions, given as follows:
\begin{equation}
    \rho_{\rm d}= \rho_{\rm Z} \, f_{\rm d/m} \, \frac{(n_{\rm HI} + 2 n_{\rm H_2}+f_{\rm ion}n_{\rm HII})}{n_{\rm H}},
\end{equation}
where $\rho_Z$ is the metal density, $f_{\rm d/m}=0.4$ is the dust-to-metal ratio
\citep[e.g.,][]{draine07}, and $n_{\rm HI}$, $n_{\rm HII}$, and $n_{\rm H_2}$ are the number density of neutral, ionized, and molecular hydrogen, respectively. We adopt $f_{\rm ion}=0.01$ for the survival probability of dust in an ionized medium following \citet{laursen09}.

Once the optical depth of each ray is measured, we combine the attenuated spectra for different directions and star particles. The galactic escape fraction ($\fesctd$) is then obtained by comparing the number of ionizing photons in the attenuated spectrum to that of the intrinsic spectrum, as follows:
\begin{equation}
 f_{\rm esc}^{\rm 3D}(t) \equiv \frac{\sum_i f_{\rm esc, i}^{\rm 3D}(t)\Dot{N}_i (t)}{\sum_i\Dot{N}_i(t)},
 \label{eq:fesc_def1}
\end{equation}
where $ f_{\rm esc,i}^{\rm 3D} (t)$ is the three-dimensional escape fraction of an $i$-th star particle at time $t$, and can be computed as
\begin{equation}
  f_{\rm esc,i}^{\rm 3D} (t) \equiv 
  \frac{\int_{\nu_0}^{\infty}\,\sum_{j}^{\rm N_{hpix}}\,e^{-\tau_{j}(\nu, t)} f_{\rm int,i}(\nu,t)\,d\nu \, /h\nu}
  {N_{\rm hpix}\int_{\nu_0}^{\infty} \, f_{\rm int,i} ({\nu,t}) \, d\nu/ \, h\nu} .
   \label{eq:fesc_def2}
\end{equation}
Here, $\nu_0$ is the frequency at the Lyman limit, and $f_{\rm int, i}$ is the intrinsic spectrum for the $i$-th star particle in units of ${\rm erg \, s^{-1} \, Hz^{-1}} $, $\rm N_{\rm hpix}$ is the total number of directions along which the escape fraction of a stellar particle is measured, and $\tau_{j}(\nu, t)$ is the total optical depth of all species for LyC photons along the $j$--th direction at time $t$.

\section{Results}

 \begin{figure*}
    \centering
    \includegraphics[width=\textwidth]{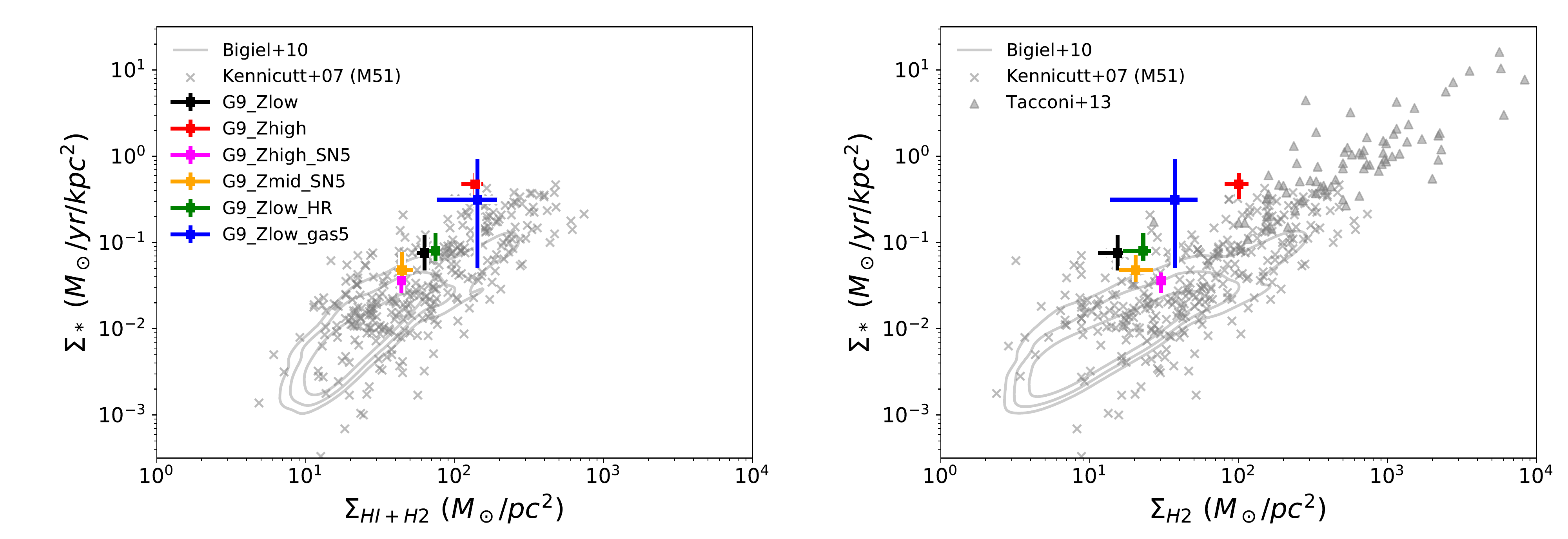}
    \caption{
     The Kennicutt-Schmidt relation for our simulations at  $150 < \tsim <300 \, {\rm Myr}$. The star formation rate surface densities ($\Sigma_{*}$) are measured within the stellar half-mass radius using the total mass of stars younger than 10 Myr. The filled squares indicate the neutral (left panel; $\rm HI$+$\rm H_{\rm 2}$) or molecular (right panel; $\rm H_{\rm 2}$) hydrogen surface density of the simulated galaxies.
    Different colour-codes correspond to the  Kennicutt-Schmidt relation from different runs, as indicated in the legend. 
    For comparison, we include the observational results by \citet{bigiel10} (grey contours representing 1, 2, and 3$\sigma$), \citet{kennicutt07} (grey x marks), and \citet{tacconi13} (grey triangles). Our simulated galaxies are consistent with the observational data, suggesting that star formation is reasonably well controlled by stellar feedback.
    }
    \label{fig:fig2}
\end{figure*}

\begin{figure*}
    \centering
    \includegraphics[width=\textwidth]{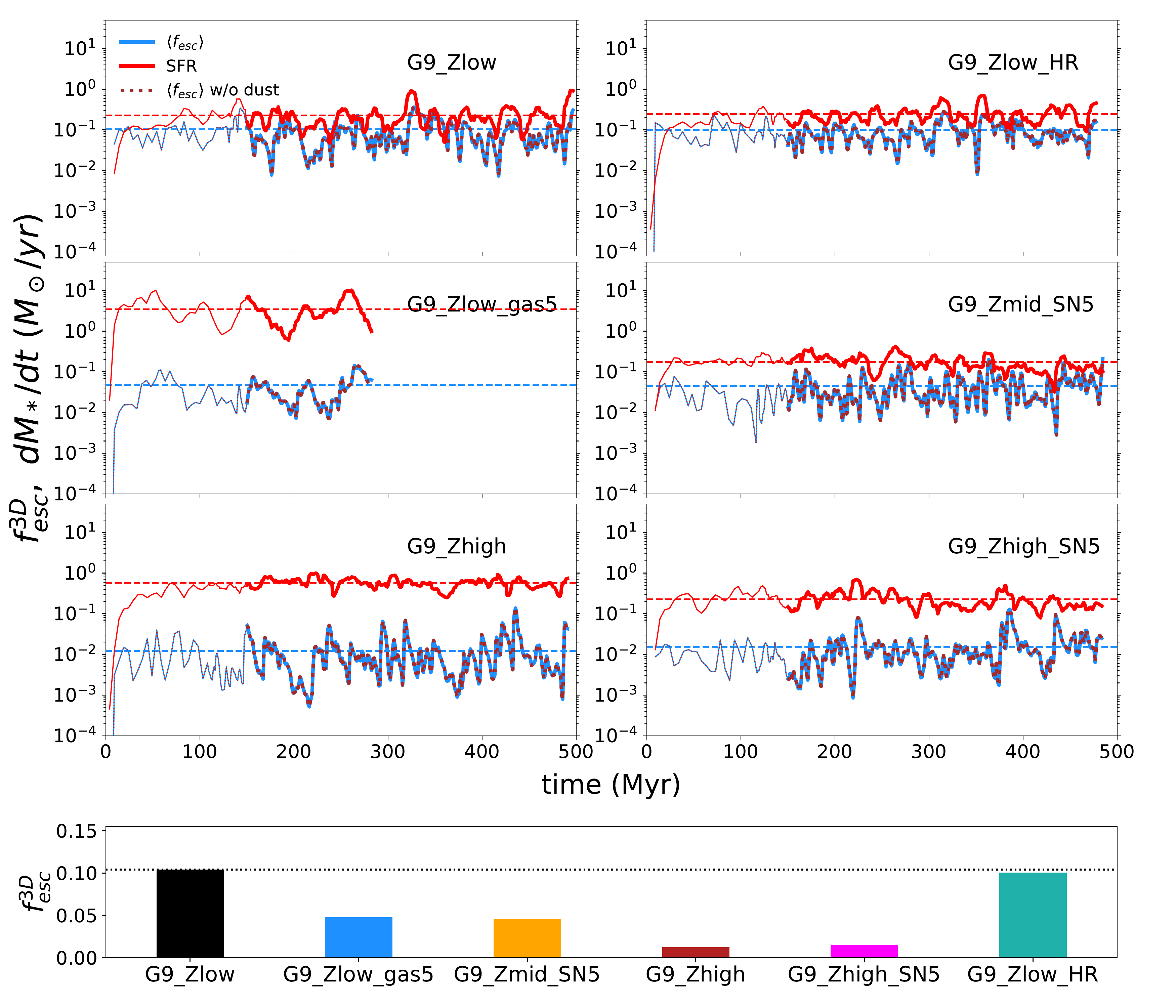}
    \caption{Upper six panels: galactic escape fractions (blue solid lines) and star formation rates averaged over 10 Myr (red solid lines) for six different runs. Thick solid lines display the epoch at which the disc appears to be more or less settled ($\tsim>150\,{\rm Myr}$).  The luminosity-weighted average escape fractions and star formation rates at $\tsim>150\,{\rm Myr}$ are shown as horizontal dashed lines.  Brown dotted lines, which are nearly identical to the blue solid lines, denote the escape fractions measured without the absorption due to dust. The escape fractions fluctuate by at least an order of magnitude with a typical time delay of $\sim 10\,{\rm Myr}$ between the peak of star formation and the peak of the escape fraction. The bottom panel displays the luminosity-weighted mean (\fesctdL) measured for $\tsim>150\,{\rm Myr}$. The escape fraction is significantly reduced when the metallicity is increased, but is fairly insensitive to the total amount of gas mass in the disc. }   
    \label{fig:fig3}
\end{figure*}
 
 In this section, we introduce the general properties of the simulated galaxies, and compare the galactic escape fractions in different runs. Based on these, we attempt to uncover what physical property is mainly responsible for the regulation of the LyC escape in massive star-forming galaxies. 

\subsection{General properties of the simulated galaxies}

Our disc galaxies start with a smooth gaseous disc, which fragments rapidly into cold dense clumps, as the pre-existing stellar disc does not provide any feedback energy. Once these clumps collapse and become gravitationally well bound, new stars form in a bursty fashion based on the local thermo-turbulent conditions (Sec. 2.1). Radiation from the stars and explosion from SNe then over-pressurise their birth clouds, sometimes driving strong outflows. When star formation occurs in the outer part of the gaseous disc, stellar feedback creates low-density holes in the ISM through which LyC photons can easily escape (see Fig~\ref{fig:fig1}). In contrast, the outburst in the central region is relatively weaker because the pressure from SNe exploding in the dense region is not very significant compared with the ambient pressure. Furthermore, even though the galactic centre is filled with young stars, their radiation is often not strong enough to ionize the whole central core ($r<1\,{\rm kpc}$) (\texttt{G9\_Zlow} in Fig.~\ref{fig:fig1}). As a result, a considerable amount of LyC photons produced from the central stars is absorbed by optically thick neutral hydrogen.
 
Fig.~\ref{fig:fig2} shows the  time-averaged  Kennicutt-Schmidt (KS) relations at $150<\tsim<300\,{\rm Myr}$ in different runs. We calculate the star formation rates by counting the total stellar mass formed within the stellar half-mass radius ($r_{\rm eff,m}$) over the past 10 Myr. The neutral ($\rm HI$+$\rm H_{\rm 2}$) or molecular ($\rm H_{\rm 2}$) hydrogen surface density are computed within $r_{\rm eff,m}$. For comparison, we also display the observed  Kennicutt-Schmidt relations in the local Universe \citep{kennicutt07,bigiel10} and at high redshifts ($z\sim1$--$3$) \citet{tacconi13}.
We find that our simulated galaxies are in reasonable agreement with the observations. The star formation rate surface densities in the runs with the normal feedback (\texttt{G9\_Zlow} and \texttt{G9\_Zhigh}) appear somewhat higher than those from the local galaxies (right panel), but these are still consistent with the trend from galaxies at high redshift \citep{tacconi13}. The runs with boosted SN feedback exhibit properties that are more in line with the local galaxies, as star formation is regulated more efficiently for a given gas surface density.

Compared with the previous study adopting the same initial conditions \citep{rosdahl17}, the average star formation rates are significantly reduced from $\sim 0.8 \, \msunyr$ \citep[see Fig.~2 of][]{rosdahl17} to $\sim 0.2\,\msunyr$, indicating that star formation is well controlled. This can be attributed to the fact that stars form in a more bursty fashion than the simple density-based star formation recipe used in \citet{rosdahl17} and that stellar feedback becomes more coherent in space and time. Moreover, extra pressure from resonantly scattered Lyman alpha photons included in this study can further suppress star formation \citep{kimm18}.

The typical half-mass radius in the gaseous disc is $\sim 1 \,{\rm kpc}$, which is similar to the typical size of LAEs \citep[e.g.,][]{gawiser07}. As seen in Fig.~\ref{fig:fig3}, the star formation rate in the runs with the fiducial gas fraction ranges from $ 0.04\, \msunyr$ to $ 1.0 \,\msunyr$, whereas it is a factor of $\sim 10$ larger on average in the gas-rich run ($ 3.5\,\msunyr$).  We note that our fiducial case forms fewer stars compared to the typical LBGs ($\ga 10\,\msunyr$), which is likely due to the fact that our simulated halo is smaller than the typical host halo mass of of LBGs. But because the primary goal of this study is to investigate the physical origin of the low escape fraction, we will continue our discussion bearing this difference in mind.

\subsection{Overview of LyC escape}

We now present the galactic average of the escape fractions from different runs and discuss how the escape fractions vary from clump scales to galactic scales.
 
\subsubsection{Galactic averaged escape fraction}

Fig.~\ref{fig:fig3} shows the galactic escape fractions and star formation rates averaged over all stellar particles within a galaxy as a function of time. As shown in previous studies \citep[e.g.,][]{wise09,kim13b,kimm14, trebitsch17}, escape fractions fluctuate as much as 1.0--1.5 dex on a timescale of $10\, \la \Delta t \la 50\,{\rm Myr}$. The fluctuating behaviour is also seen in star formation rates, but with the offset of $\Delta t\sim5$--$20\, {\rm Myr}$  from that of the escape fractions. The asynchronous correlation can be interpreted as the feedback cycle, which is set by the formation of young stars in a dense gas clump and the subsequent destruction due to stellar feedback \citep{kimm14}. We will discuss this in more detail in the next section.

We find that the luminosity-weighted average of the galactic escape fraction after the galaxy becomes settled, i.e. $150 < \tsim < 500\,{\rm Myr}$, is $\fesctdL=10.4\,\%$ in the fiducial run (\texttt{G9\_Zlow}).
If we increase the metallicity to $Z=0.02$, the average escape fraction is very significantly reduced to $\fesctdL \sim1\%$, regardless of whether SN feedback is boosted (\texttt{G9\_Zhigh\_SN5}) or not (\texttt{G9\_Zhigh}).
 When the typical metallicity of LBGs is used \citep[$Z=0.006$, e.g.,][]{mannucci09,onodera16} with boosted SN feedback (\texttt{G9\_Zmid\_SN5}), the escape faction is decreased by a factor of two to $\fesctdL\approx 5\%$.
This indicates that metallicity plays a significant role in determining the escape fraction of LyC photons. 

Along with metallicity, the amount of gas in a galaxy can affect the escape fraction, as the absorption of LyC radiation depends mostly on the column density of neutral hydrogen. However, it is not obvious whether the increased gas mass simply lowers the escape fraction because it may lead to an opposite trend by enhancing the star formation and hence the stellar feedback. The comparison between the fiducial and gas-rich run shows that the former effect is more dominant. The escape fraction in the \texttt{G9\_Zlow\_gas5} run, where the amount of gas mass is augmented by a factor of 5, is reduced to $\fesctdL= 4.8 \,\%$, even though the star formation rates are enhanced by a factor of $\sim 15$.

Of the five species considered for the photo-absorption in our simulations, the primary agent responsible for LyC absorption is neutral hydrogen. The contribution from molecular hydrogen and helium is minor because the effective optical depth  ($\tau_{\rm eff,X}\equiv - \ln \left< f_{\rm esc,X}^{\rm 3D} \right>_\mathcal{L}$) is low ($\tau_{\rm eff, H2}= 0.59$, $\tau_{\rm eff, He}=0.28$). We find that absorption due to dust is also negligible (Fig.~ \ref{fig:fig3}, dotted lines; $\tau_{\rm eff, dust}= 0.87$\footnote{Although the effective optical depth due to dust seems to be rather high, their actual contribution to the absorption of LyC photons is negligible, as neutral hydrogen preferentially absorbs them  in regions where the dust optical depth is significant. }). The difference in the escape fraction with and without dust is in the order of $\sim10^{-3}$--$10^{-4}$. This is mainly due to our assumption that only 1\% of dust can survive in the ionized medium.  If we assume that dust can survive even at high temperatures, the escape fraction would be reduced by $\sim  17 \,\%$ (from $\fesctdL= 10.4 \,\%$ to  $ 8.6\,\%$) in the fiducial run, and a higher fraction of LyC photons ($\sim  37\,\%$ would be absorbed by dust in the metal-rich (\texttt{G9\_Zhigh}) run (from $\fesctdL= 1.2 \,\%$ to $ 0.7\,\%$).

\begin{table}
    \centering
    \begin{tabular}{l c c c}
    \hline 
    Simulations&  $\langle \fesctd \rangle_{\mathcal{L}}$&$\langle f_{\rm esc, nodust}^{\rm 3D}\rangle_{\mathcal{L}}$ & $\langle dM_*/dt \rangle$   \\
     & (1) & (2) & (3) \\
    \hline
    
    \texttt{G9\_Zlow} &  0.1041 & 0.1044 & 0.2279 \\
    \texttt{G9\_Zhigh}  & 0.0123 & 0.0123 &   0.5784 \\
    \texttt{G9\_Zlow\_gas5}  & 0.0479 & 0.0481 & 3.4798 \\
    \texttt{G9\_Zmid\_SN5}   & 0.0464 & 0.0465 & 0.1734 \\
    \texttt{G9\_Zhigh\_SN5}   & 0.0152& 0.0153 & 0.2280 \\
    \texttt{G9\_Zlow\_HR} &  0.1005 & 0.1007 &  0.2457   \\

    \hline
    \end{tabular}
    \caption{Time-averaged galactic escape fraction of LyC photons and star formation rates: (1) luminosity-weighted average of the escape fraction at $\tsim>150\,{\rm Myr}$,
    (2) average without the absorption due to dust, and (3) star formation rates averaged over $\tsim>150\,{\rm Myr}$.}
    \label{tab:fesc}
\end{table}

\subsubsection{The escape fraction of young stars in gas clumps}

Most of the LyC radiation re-ionizing the Universe is thought to arise from stars younger than $\sim 10~{\rm Myr}$  \citep[e.g.,][]{leitherer99,stanway16}. However, sites of star formation are very dense ($n_{\rm H}>10^2$ cm$^{-3}$), and most emitted flux is likely to be absorbed by the birth cloud \citep{kimm19,kimj19}. These dense clumps are often suspected to be the main causes of our low escape fractions  \citep{dove94,yajima11,kim13b,paardekooper15,ma15,kimm17}. Eventually, the clumps should be disrupted to produce large-scale outflows, as commonly observed in star-forming galaxies \citep{steidel10}. 
However, the correlation between the timescale for the dispersal of the dense clump and the  {\it galactic} escape fraction is not discussed in detail in radiation-hydrodynamics simulations \citep[c.f.,][]{dale13,howard18,kimm19,kimj19,kakiichi19}.

\begin{figure}
    \centering
    \includegraphics[width=8.3cm]{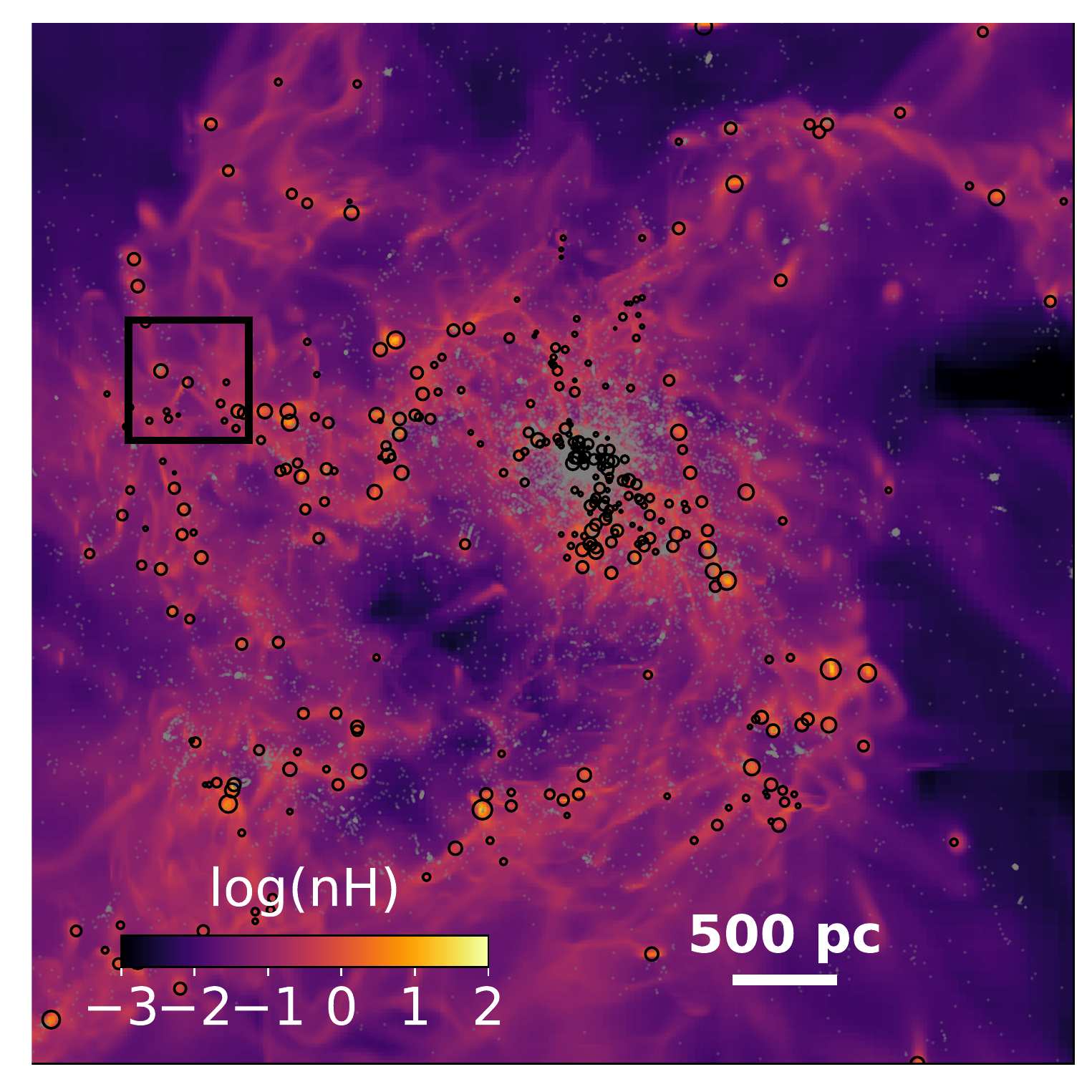}
    \caption{Gas clumps (black circles) identified with the clump finder \texttt{PHEW}. The background image shows the projected gas density in the \texttt{G9\_Zlow} run at $\tsim= 160\,{\rm Myr}$. The width of the image is $\approx$ 3 kpc. One can see that the gas clumps are preferentially formed in the dense region of the filamentary structures in the ISM. The black empty square shows an example of a patch of the ISM where star formation occurs in a bursty fashion and the escape fraction becomes very high eventually. }
    \label{fig:fig4}
\end{figure}

\begin{figure*}
    \centering
    \includegraphics[width=\textwidth]{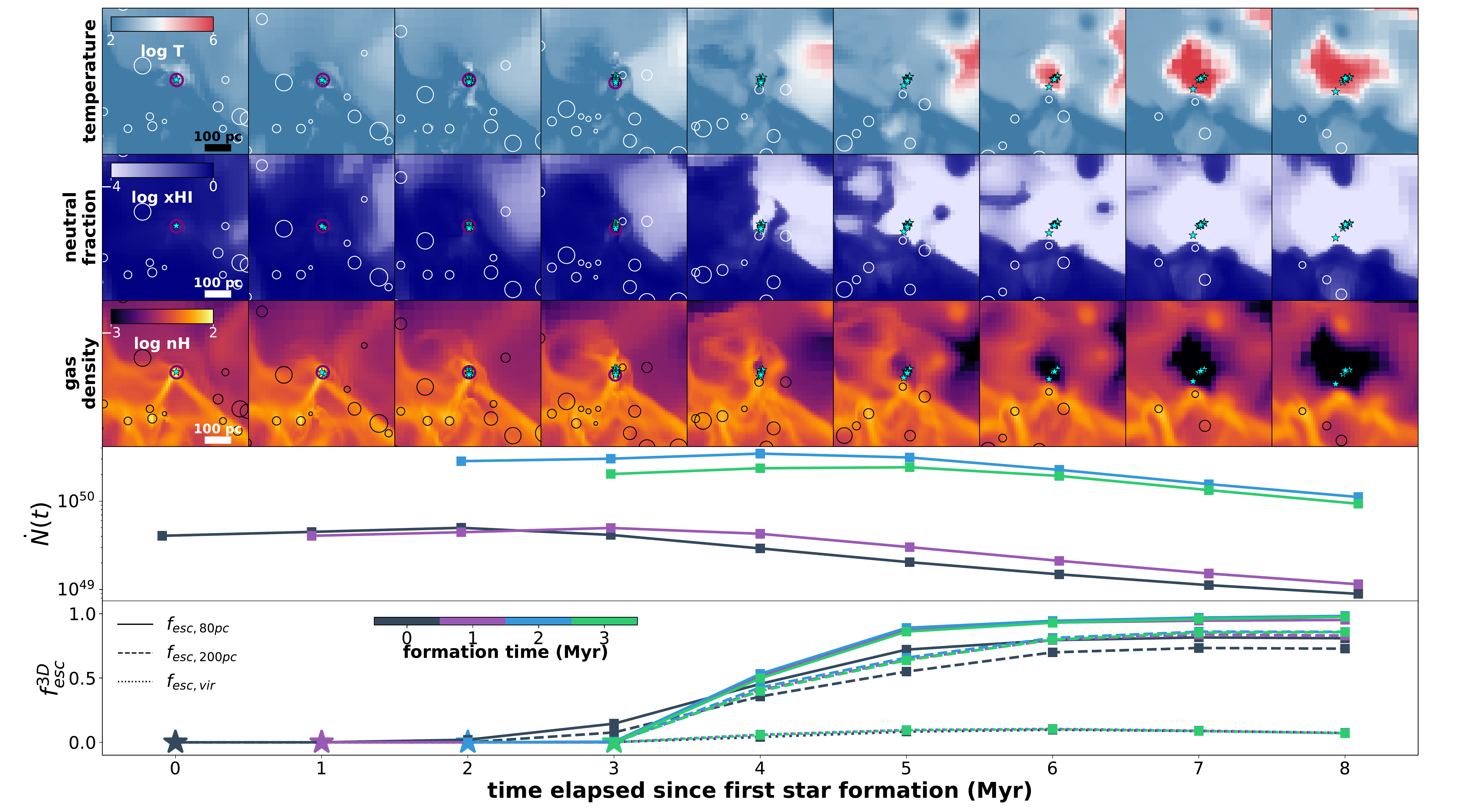}
    \caption{ Evolution of gas clumps marked in Fig~\ref{fig:fig4} inside black empty squares. The three upper rows show zoomed-in distributions of temperature, neutral hydrogen fraction, and gas density. The main clump is shown as the purple circle, and the newborn star particles in the clump are indicated as cyan stars. The luminosity-weighted escape fraction (bottom) and subtotal of the ionizing emissivity (second to the bottom) of these stars are coloured based on their formation time. In the bottom panel, different line styles correspond to the escape fraction measured at a distance of 80 pc (solid), 200 pc (dashed), or the virial radius (dotted) from the position of each star particle. This shows an example in which the escape fraction increases with time as stellar feedback disrupts the clump.}
    \label{fig:fig5}
\end{figure*}

To give some clarity to our understanding of how LyC radiation escapes from galaxies, we identify gas clumps in the simulated galaxies using the 3D clump-finding algorithm \textsc{PHEW} \citep[Parallel HiErarchical Watershed,][]{bleuler15} implemented in \textsc{Ramses}. This algorithm dissects the region around density peaks above a certain threshold along the watershed and eliminates the noise with which the peak-to-saddle ratio (relevance) becomes smaller than some threshold. Following \citet{grisdale18}, we use a density threshold of $\rho_{\rm thres}=100\,{\rm cm^{-3}}$, relevance threshold of $r=1.2$, and saddle threshold of $\rho_{\rm saddle}=10^4\,{\rm cm^{-3}}$. Note that \textsc{PHEW} does not assume any shape in identifying structures, but we refer to a sphere corresponding to the total volume of the cloud to indicate the size. We also note that since our resolution is limited ($\Delta x_{\rm min} \sim 10\,{\rm pc}$), our clumps do not necessarily represent observed GMCs with turbulent structures but rather represent candidates of stellar nurseries and immediate absorbers of LyC photons. 
We attempt here to understand how the general properties of the star-forming clumps affect the escape fractions in radiation-hydrodynamics simulations.

Fig.~\ref{fig:fig4} presents the typical distribution of clumps in the fiducial run where the dense clumps are formed along filamentary structures. The typical radius of the clumps is $R_{\rm clump}\sim 30\,{\rm pc}$, and their gas mass ranges from $10^5$ to $10^7\,\msun$ with an average of $10^6\,\msun$ in the case of the \texttt{G9\_Zlow} run, which is broadly consistent with observations \citep[e.g.,][]{colombo14}.  We find that over 90\% of  newborn stars ($\tage<1 \, {\rm Myr}$) are still embedded in the gas clumps identified by \textsc{PHEW}, whereas the rest ($\la 10\%$) reside in the ISM mainly because radiation feedback operates early and destroys the clouds. 

The effect of feedback on cloud evolution is more clearly illustrated in Fig.~\ref{fig:fig5}, which corresponds to a small patch of the ISM hosting a typical cloud with a bursty star formation event, shown as the black box in Fig.~\ref{fig:fig4}. The projected distributions of density, neutral hydrogen fraction, and temperature show that the dense pocket of the collapsing cloud initially forms stars, and then radiation pressure and photo-ionization heating blow the gas away, creating low-density channels. Once SNe start to explode 4 Myr after their birth, hot gas with $T\ga10^7 \, {\rm K}$ develops around the cluster (see the first row of Fig.~\ref{fig:fig5}). The escape fraction measured at 80 pc, which is just outside the clump, reaches nearly 100\% after 5 Myr, whereas the fraction measured at 200 pc increases more slowly. The instantaneous escape fractions measured at the virial radius ($R_{\rm vir}=89\,{\rm kpc}$) rise only up to  $\sim7\%$ because a large fraction of the photons are absorbed by neutral interstellar gas outside the clump.

\begin{figure}
    \centering
    \includegraphics[width=8.3cm]{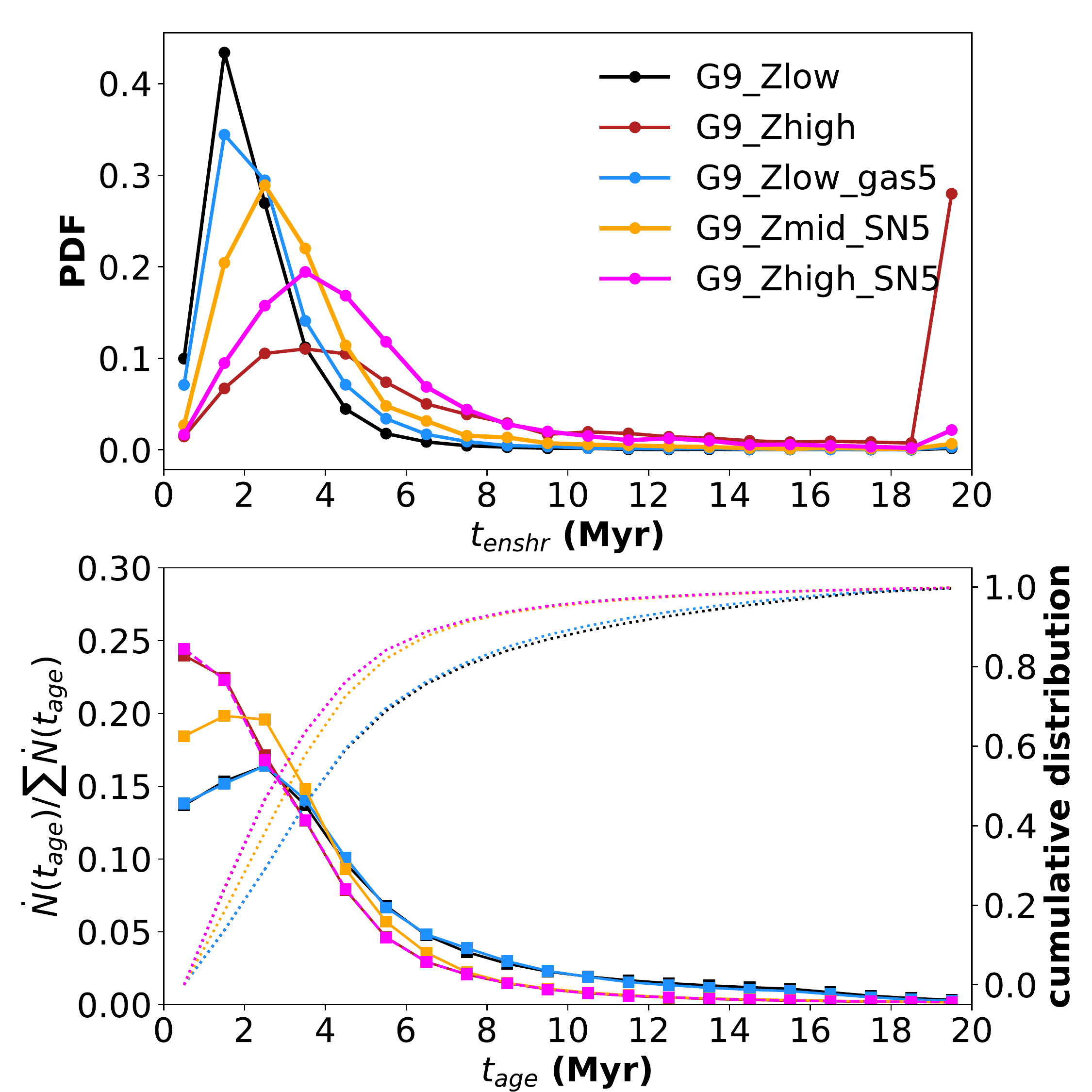}
    \caption{Upper panel: the probability density distributions (PDFs) of the enshrouded timescale (\tensh) of star particles. We trace all newborn stars from $\tsim\sim150$ Myr to $\sim500\,{\rm Myr}$ (to $\sim300\,{\rm Myr}$ for \texttt{G9\_Zlow\_gas5}) in the simulation until they decouple from any gas clumps.  Star particles that never leave the gas clumps within 20 Myr from birth are all shown as $\tensh=20\,{\rm Myr}$.  Lower panel: the fractional ionizing emissivity (solid lines, $\dot{N}(\tage)/\sum\dot{N}(\tage)$) of stars with different ages at $\tsim> 150\,{\rm Myr}$. The fractional ionizing emissivity from \texttt{G9\_Zlow} is very similar to that from \texttt{G9\_Zlow\_gas5} because the input stellar SEDs are the same. The dotted lines indicate the cumulative contribution ($\Sigma \dot{N}(\tage)$). Newborn stars in the higher metallicity runs are enshrouded by the dense gas clumps for a longer period of time. The fractional ionizing emissivity from stars younger than 2 Myr is also higher in more metal-rich runs.}
    \label{fig:t_trap}
\end{figure}

\begin{figure}
    \centering
    \includegraphics[width=8.5cm]{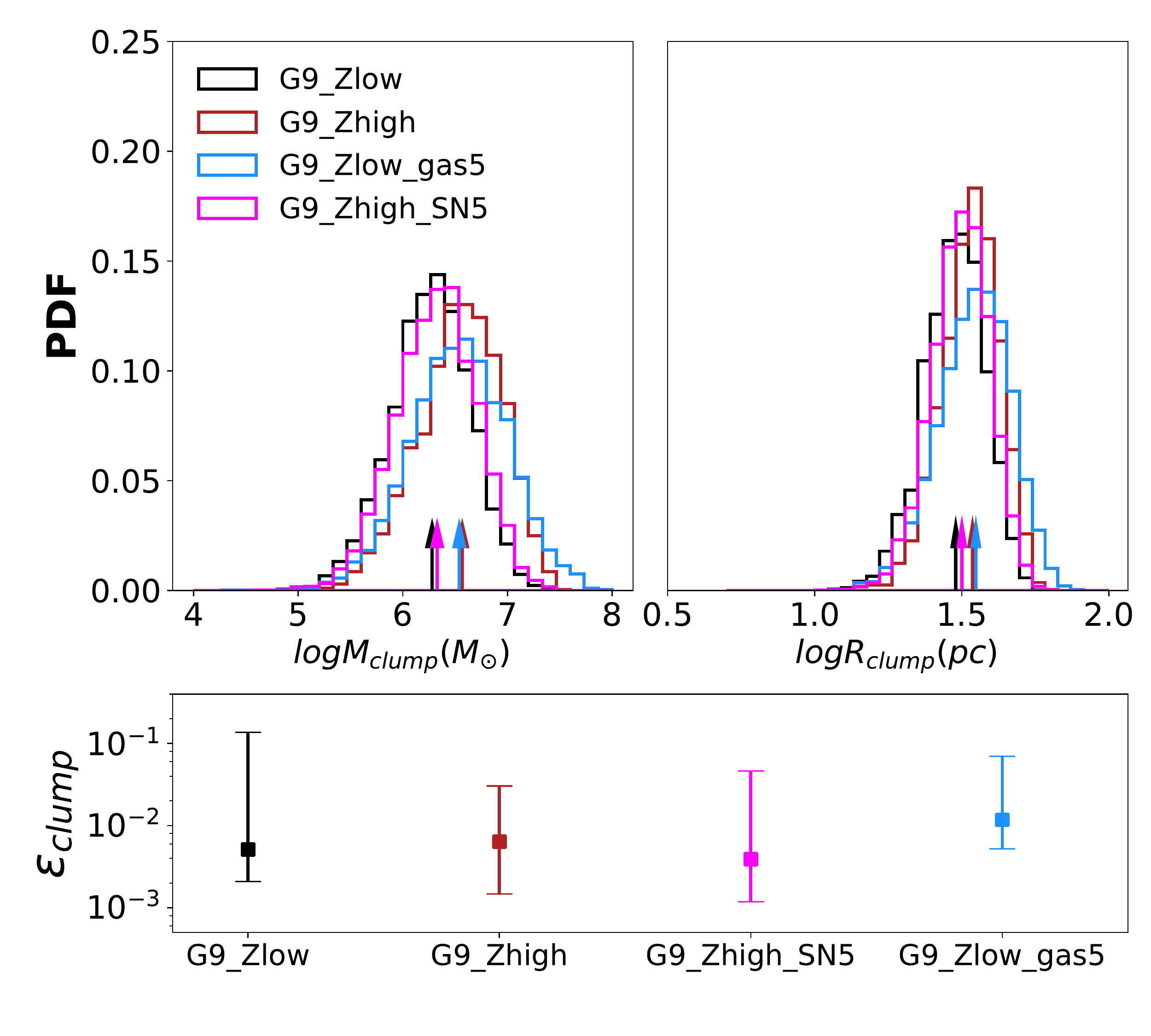}
    \caption{Upper panel: size and mass distribution of the star-forming gas clumps at $\tsim> 150\,{\rm Myr}$. The arrows in each panel indicate the median radius and mass of the star-forming clumps. Different colour-codes represent the results from different runs, as shown in the legend. Bottom panel: we show the total mass of stars formed in the last 1 Myr divided by the total mass of star-forming clouds ($\epsilon_{\rm clump}\equiv M_{\rm star}(<1\,{\rm Myr}) / (M_{\rm star}(<1\,{\rm Myr})+M_{\rm clump})$) as a rough estimate of the instantaneous star formation efficiency. The error bars indicate the interquartile range of the distributions with the median value. The properties of the gas clumps in the simulated galaxies are reasonably similar, although minor differences can be found (see the text).    }
    \label{fig:cloud_prop}
\end{figure}

In Fig.~\ref{fig:t_trap}, we measure how long each newborn star particle is enshrouded by a gas clump (\tensh). To do so, we trace newborn star particles between  $150$ and $500\,{\rm Myr}$ of the simulation run-time ($300\,{\rm Myr}$ for \texttt{G9\_Zlow\_gas5}) until they are detached from the boundary of their host clump defined as a sphere. The top panel demonstrates that more than 80--90\% of the stars are no longer associated with their birth place within a short timescale ($\tensh \la \,5 \,{\rm Myr}$) in the fiducial run ( black lines). This is consistent with the claim that the observed gas clouds are rapidly dispersed on $\sim 5 \, {\rm Myr}$ by stellar feedback \citep[e.g.,][c.f. \citealt{engargiola03,kawamura09}]{hartmann01,kruijssen19}. In particular, we find that $\sim 80\%$ of the stars younger than $\approx 3 \, {\rm Myr}$ are not enshrouded by their birth cloud, suggesting that photo-ionization heating and Ly$\alpha$ radiation pressure are primarily responsible for this early escape.  In principle, some of the first star particles formed in a clump could explode as SNe and help to blow away the clump, lowering \tensh\ of the star particles that are formed late.  However, we confirm that the average \tensh\ of the star particles formed for the first time in a clump between two consecutive snapshots is only  2.3 Myr in the fiducial run, which is still shorter than the typical lifetime of massive stars. This is of course not true in the metal-rich run where a significant fraction of stars is trapped for $\ga 4 \, {\rm Myr}$. In this case, it would be more reasonable to conclude that not only radiation feedback but also early SNe are responsible for the disruption of the cloud.

\subsection{Effects of gas metallicity on the escape fraction}

Recently, \citet{kimm19} demonstrated that the LyC escape fraction increases with decreasing cloud mass and increasing star formation efficiency \citep[see also][]{kimj19}. To identify the physical origin of the metallicity dependence, we examine how metallicity affects the properties of gas clouds, such as cloud mass and radius, in the upper panel of Fig.~\ref{fig:cloud_prop}. We find that the median clump masses in the \texttt{G9\_Zlow} and \texttt{G9\_Zhigh} run are $M_{\rm clump}= 1.9 \times10^6\,\msun$ and $ 3.0\times10^6\,\msun$, respectively, and the median radii of the clouds are  $R_{\rm clump}= 30\,{\rm pc}$ and $34\,{\rm pc}$, respectively. Even when the extreme SN feedback model is adopted (\texttt{G9\_Zhigh\_SN5}), the clumps in the metal-rich run turn out to show similar mass and radius distributions ($M_{\rm clump}= 2.1\times10^6\,\msun$, $R_{\rm clump}=32\,{\rm pc}$). Given that the average escape fractions in both metal-rich runs are lower ($\fesctd\approx1\%$) than the metal-poor case ($\fesctd\approx 10\,\%$), the cloud properties in different metallicity runs are unlikely to be responsible for different $\fesctd$.

We also examine the {\em instantaneous} star formation efficiency of star-forming clumps ($\epsilon_{\rm clump}$) in the lower panel of Fig.~\ref{fig:cloud_prop}. We define $\epsilon_{\rm clump}$ as the total mass of stars that are formed in the last 1 Myr and that are still embedded in a clump, divided by the sum of newly formed stellar mass and  clump mass at each snapshot.  Here 1 Myr is chosen to roughly measure the burstiness of star formation, which is shown to be important in determining the escape fraction \citep[][]{dale13,kimm17,kimm19}.  The metal-poor disc (\texttt{G9\_Zlow}) shows  a median $\epsilon_{\rm clump}$ of $0.5\,\%$, whereas $\epsilon_{\rm clump}$ is  slightly increased to $0.6\,\%$ in the \texttt{G9\_Zhigh} run. In contrast, the value of $\epsilon_{\rm clump}$ is reduced to $0.4\%$ if strong SN feedback is used (\texttt{G9\_Zhigh\_SN5}). Although the fiducial run with the low metallicity shows a slightly lower $\epsilon_{\rm clump}$ than \texttt{G9\_Zhigh}, a larger fraction of LyC photons escapes from the metal-poor galaxy.  Thus we are led to conclude that slightly different SFEs found in the different metallicity runs do not affect the escape of LyC photons significantly.

The lower escape fractions in the \texttt{G9\_Zhigh} runs are best explained by the slow disruption of metal-rich star-forming clouds. The upper panel of Fig.~\ref{fig:t_trap} shows that the  enshrouded timescale of young stars is longer in metal-rich environments. On an average, it takes 9.8 Myr for young stars to be unveiled from their metal-rich birth cloud (\texttt{G9\_Zhigh}), whereas the mean enshrouded timescale in the \texttt{G9\_Zlow} run is 2.3 Myr.  We find that \tensh\ in the metal-rich disc is still large even when we boost SN feedback ($\tensh= 5.2 \, {\rm Myr}$). For comparison, the run with $Z=0.006$ shows an inter-mediate \tensh\ of $ 3.5\, {\rm Myr}$.   

\begin{figure}
    \centering
    \includegraphics[width=8.3cm]{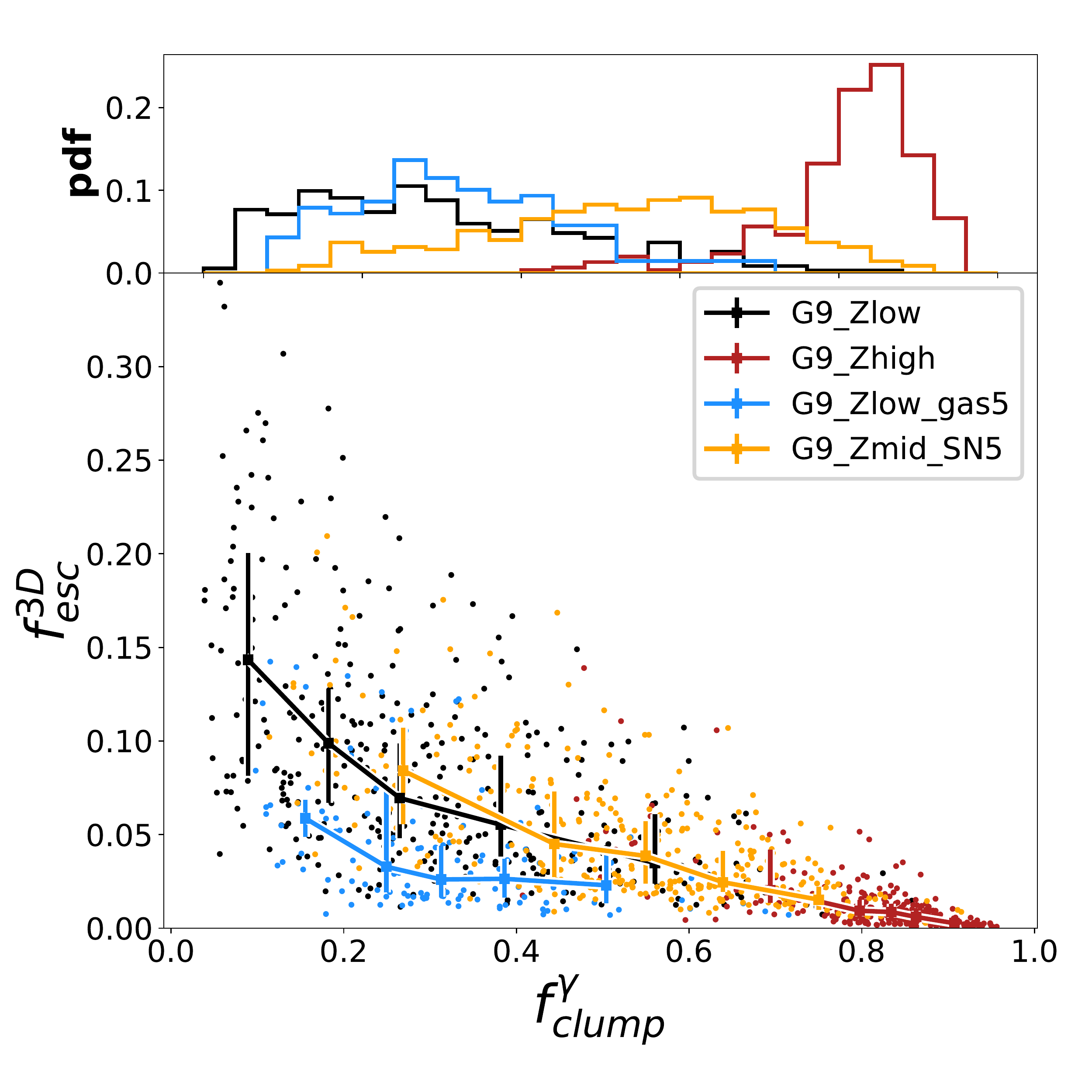}
    \caption{
      The fraction of LyC photons produced by stars embedded in gas clumps to the total ionizing radiation ($f_{\rm clump}^{\gamma}$). The bottom panel shows the relation between the escape fraction and the fraction of ionizing radiation emitted from within clumps. Each data point represents the luminosity-weighted LyC escape fraction and $f_{\rm clump}^{\gamma}$ in each snapshot at $\tsim \simeq 150\,{\rm Myr}$. The error bars indicate the interquartile range. We also present the probability distribution function of $f_{\rm clump}^{\gamma}$ in different runs in the upper panel.  It can be seen that the escape fraction is lower when more of the radiation is produced inside gas clumps. 
}
    \label{fig:fclump}
\end{figure}

Fig.~\ref{fig:fclump} further corroborates that the escape fraction is closely linked to the fraction of young stars embedded in their birth clouds. We calculate the ratio of the number of intrinsic LyC photons produced by the stars located inside clumps and the total number of intrinsic ionizing photons from the entire galaxy, as $\fclump \equiv \dot{N}_{\rm clump} / \dot{N}_{\rm tot}$. Each point in Fig.~\ref{fig:fclump} denotes the luminosity-weighted average of instantaneous \fesctd\ and \fclump\ in each snapshot. There is a clear trend that \fesctd\ decreases with increasing \fclump. In particular, it can be seen that a large fraction ($50$--$90\%$) of ionizing radiation is produced inside birth clouds in metal-rich galaxies (\texttt{G9\_Zhigh}), whereas the fraction is significantly smaller ($\fclump\sim10$--$50\%$) in metal-poor discs (\texttt{G9\_Zlow}). This is consistent with the findings from the high-resolution ($\Delta x_{\rm min}=0.25\,{\rm pc}$), GMC simulations conducted by \citet{kimm19}. 

The dependence of \tensh\ and \fclump\ on metallicity can be attributed to several factors. First, more metal-rich stellar populations emit fewer ionizing photons \citep{leitherer99}, leading to less significant radiation feedback. Second, radiative metal cooling becomes enhanced in the metal-rich cases, lowering the thermal pressure in the vicinity of young stars. Finally, Ly$\alpha$ pressure becomes weaker in the metal-rich medium because photons are more likely to be destroyed by dust before they impart radial momentum to the surroundings. For example, \citet{kimm18} show that the maximum multiplication factor of Ly$\alpha$ photons is $M_{\rm F}\sim 120$ at $Z=0.1\,Z_\odot$, whereas the maximum $M_F$ at $Z=Z_\odot$ is $\sim 50$, assuming the metallicity-dependent dust-to-metal ratio derived from \citet{remyruyer14}. As a result, disruption takes place more slowly, and the young stars are enshrouded for a longer time.

The metallicity has a strong effect on the time-evolution of the SED \citep[see e.g. ][]{rosdahl18} which in turn has a significant effect on the escape fraction. We calculate the ionizing emissivity as a function of stellar age ($\dot{N}(\tage)$) and measure the contribution of each stellar population (with different ages) to the total number of ionizing photons. This fractional ionizing emissivity ($\dot{N}(\tage)/\sum\dot{N}(\tage)$) is averaged over the period from $\tsim=150$ to $500\,{\rm Myr}$ (except \texttt{G9\_Zlow\_gas5} where we take the average between $\tsim=150$ and $300\,{\rm Myr}$). It is clear from the Fig.~\ref{fig:t_trap} that more metal-rich populations produce a larger fraction of ionizing radiation at $\tage\la 5\,{\rm Myr}$. This means that a smaller fraction of the total ionizing radiation would escape in the metal-rich case than the metal-poor ones, even if their enshrouded timescales are equally short ($\tensh\sim 5 \, {\rm Myr}$).

\begin{figure}
    \centering
    \includegraphics[width=8.3cm]{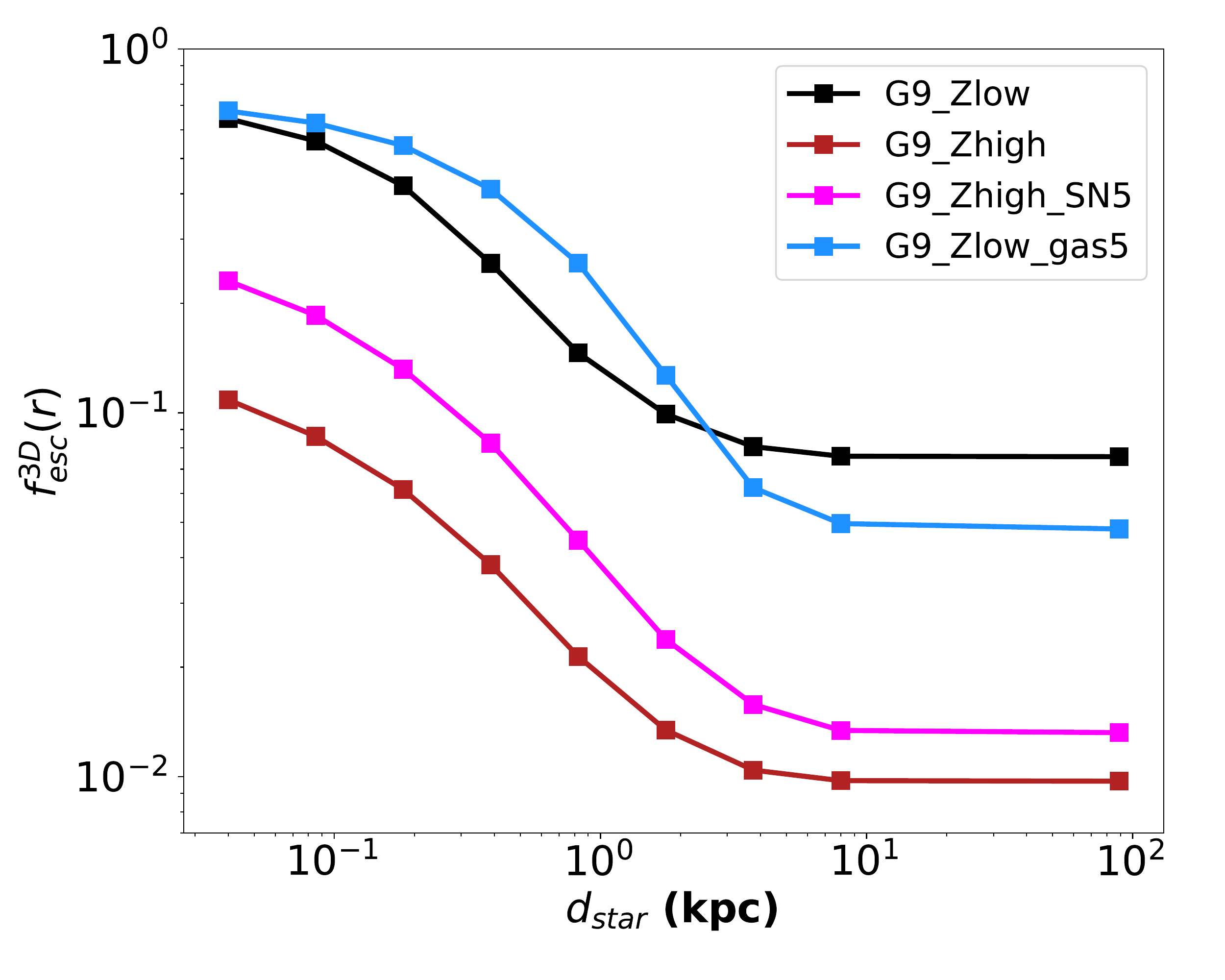}
    \caption{The luminosity-weighted escape fraction measured at different distances from each star particle ($\fesctd(r)$) by stacking the data at $150\,{\rm Myr}<\tsim<300\,{\rm Myr}$. Different colour-codes correspond to different simulations. The escape fractions measured on clump scales ($ 40 \,{\rm pc}$) are similar between \texttt{G9\_Zlow} and \texttt{G9\_Zlow\_gas5}, but $\fesctd\,(r)$ at large distances becomes lower in the \texttt{G9\_Zlow\_gas5} run, as the extended gaseous disc absorbs more LyC photons. Note that the metal-rich runs show systematically lower $\fesctd(r)$ because a large amount of ionizing radiation is immediately absorbed by their birth clumps. }
    \label{fig:fesc_distance}
\end{figure}

\subsection{Effects of gas mass on the escape fraction}

We now look at how the gas mass, or the gas fraction, of the galaxy may affect the escape fraction. The upper panel of Fig.~\ref{fig:cloud_prop} shows that the clumps in the gas-rich run (\texttt{G9\_Zlow\_gas5}) are slightly bigger ($R_{\rm cloud}=35\,{\rm pc}$) and more massive ($M_{\rm cloud}= 3.5\times10^6\,\msun$) than the fiducial case (\texttt{G9\_Zlow}). The instantaneous star formation efficiency is also enhanced from $\epsilon_{\rm clump}= 0.5\%$ to $ 1.2\%$ on average. However, the increase in gas mass by a factor of five does not make a significant difference in the enshrouded timescale of young star particles (Fig.~\ref{fig:t_trap}) or the fraction of ionizing radiation produced inside star-forming regions (Fig.~\ref{fig:fclump}). This suggests that the disruption of clumps is not significantly affected by the slight changes in cloud properties in our simulations.

To understand the cause of the lower \fesctdL\ of $\approx 5\%$ in the gas-rich case, we measure the escape fraction not only at the virial radius but also at various distances from each star particle. Fig.~\ref{fig:fesc_distance} shows that roughly $30\%$ of LyC photons are absorbed at the clump scale (40 pc), whereas the other $\sim 60\%$ of photons are absorbed by the ISM ($r\la 2 \,{\rm kpc}$) not only in the gas-rich run but also in the fiducial run. This demonstrates that {\em the absorption due to the ISM is about equally important as that due to the cloud in the metal-poor environments}. The absorbed fraction on cloud scales is substantially increased to $70$--$90\%$ in the metal-rich runs, again because the clouds are disrupted more slowly.

\begin{figure}
    \centering
    \includegraphics[width=8.5cm]{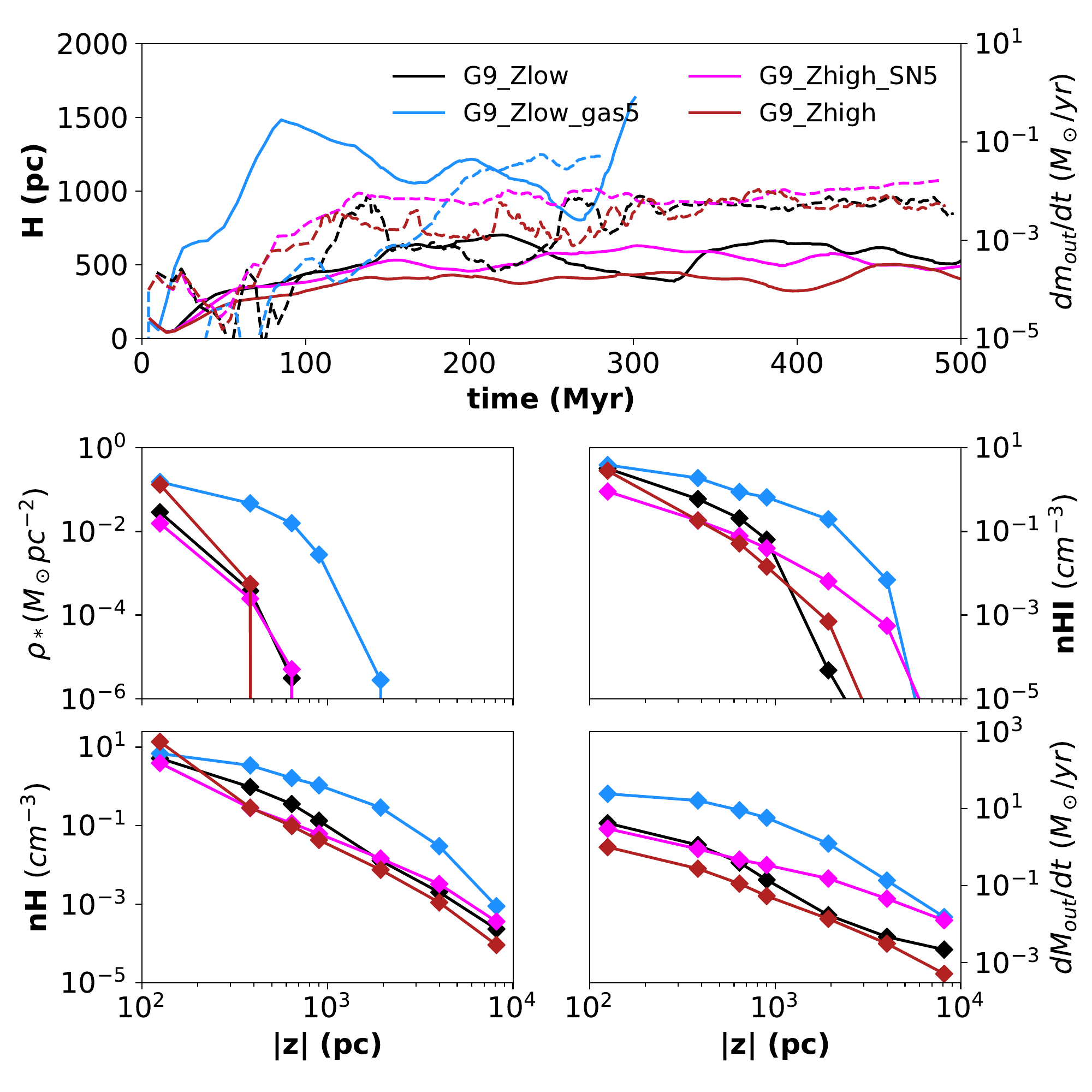}
    \caption{
    The top panel shows the scale height (solid lines) defined as the $H_{\rm gas} \equiv \sqrt{ \int \rho z^2 dV / \int \rho dV }$ and the outflow (dashed lines) of the gaseous disc,  as a function of time. Different colour-codes indicate the scale height and outflows in different runs, as indicated in the legend. The lower panels show the stellar density ($\rho_\star$), neutral hydrogen number density ($n_{\rm HI}$), outflow rates, and the total hydrogen density ($n_{\rm H}$) measured within the cylindrical volume of radius $r_{\rm eff,star}$ and the height of 10 kpc, averaged over $150< \tsim < 300\,{\rm Myr}$ for the \texttt{G9\_Zlow\_gas5} run or $150< \tsim < 500\,{\rm Myr}$ for other models. Each point indicates the average quantity measured at cylindrical bin with a width of 250 pc and different heights ($|z|$) from disc plane. One can see that the neutral gaseous disc becomes thicker and more extended if star formation activities are enhanced by increasing gas mass  (\texttt{G9\_Zlow\_gas5}) or extended if SN feedback is boosted (\texttt{G9\_Zhigh\_SN5}).}

    \label{fig:vert}
\end{figure}

Fig.~\ref{fig:vert} illustrates that the large optical depth at $r\ga 1\,{\rm kpc}$ in the gas-rich run is essentially due to the more extended gaseous disc. We measure the thickness of the gas disc to be $H_{\rm gas}\equiv \sqrt{\int \rho z^2 dVb/\int \rho dV}$. \texttt{G9\_Zlow} exhibits a typical scale height of $\sim 500 \, {\rm pc}$, whereas the \texttt{G9\_Zlow\_gas5} run shows $H_{\rm gas}\sim 1\, {\rm kpc}$. This is because more enhanced star formation builds up the mid-plane pressure and drives strong outflows, thickening the disc \citep[e.g.,][]{kim13}. As a result, the distribution of neutral hydrogen at $|z|\ga  1\,{\rm kpc}$, measured within the stellar effective radius, also becomes  extended, thus lowering the escape probability of LyC photons.

It is interesting that $\fesctd(r)$ at $r\la 1\,{\rm kpc}$ in the gas-rich run is larger than that in the fiducial case (Fig.~\ref{fig:fesc_distance}), despite the neutral hydrogen density being systemically higher (Fig.~\ref{fig:vert}). This can be attributed to the fact that more active star formation leads to a more porous ISM in the gas-rich disc. Indeed, we find that the volume filling fraction of neutral hydrogen with low density ($n_{\rm HI}\la10^{-5}\,{\rm cm^{-3}}$) at $|z|<1\,{\rm kpc}$, which allows LyC photons to escape through low-density holes easily \citep[e.g.,][]{cen15}, is higher ($15$--$27\%$) in the gas-rich disc than in the fiducial disc with lower star formation rates ($9$--$15\%$). Thus we conclude that a higher fraction of LyC photons escapes on ISM scales in the gas-rich disc but the radiation eventually becomes absorbed more efficiently by the thick gaseous disc.

\subsection{Effects of strong SN feedback on the escape fraction}

We also compare the results from the metal-rich run with and without a SN boost to gauge the impact of possible over-cooling on the escape fraction. Fig.~\ref{fig:fesc_distance} shows that the escape fraction measured on cloud scales ($\sim 40\,{\rm pc}$) is higher by a factor of  2.5 in the strong feedback run than in the fiducial run. This is not surprising because the typical \tensh\  in \texttt{G9\_Zhigh\_SN5} is $\approx  5\,{\rm Myr}$, whereas \tensh\ in the \texttt{G9\_Zhigh} run is significantly larger ($\approx 10\,{\rm Myr}$)\footnote{However, we do not expect that the enshrouded timescale becomes significantly reduced in the metal-poor galaxy with strong SN feedback because \tensh\ is already shorter than the typical lifetime of massive stars ($\tage\ga 4\,{\rm Myr}$). }. In contrast, the difference in the ratio of $\fesctd(r)$ at the virial sphere ($\approx 90\,{\rm kpc}$) is reduced to a factor of 1.5 between the two runs, indicating that the absorption of LyC photons in the ISM must be more significant in the run with strong feedback.

Fig.~\ref{fig:vert} indeed demonstrates that the simulated galaxy with enhanced SN feedback has a more extended vertical profile ($|z|\ga1\,{\rm kpc}$) of neutral hydrogen than the metal-rich system without a SN boost. This is again due to the enhanced pressure originating from extra SN energy and associated gas outflows. One can also see that the central stellar density in the \texttt{G9\_Zhigh\_SN5} run is more suppressed than \texttt{G9\_Zhigh}, and that the outflow rates are higher. Thus we conclude that strong SN feedback increases the escape fraction at small scales but reduces the differences at larger scales in our metal-rich, massive disc galaxies. However, we note that this trend may not apply to less massive galaxies where the thick gaseous disk is not well developed. For example, \citet{rosdahl18} showed that the escape fraction of LyC photons increases from $\sim 2\%$ to $\sim 12\%$ in dark matter haloes of mass $\la 10^{10}\,\msun$ when the frequency of SN explosions is augmented by a factor of four.

\section{Discussion}

We now compare the simulated escape fractions with those obtained from observations or other theoretical studies. We also examine the possible correlation between the escape fractions and star formation rate surface density and gas outflow rates,  which were used to study the reionization history of the Universe in previous works. Finally, we discuss the impacts of numerical resolution on our main conclusions.

\subsection{Comparison with observations}

\begin{figure*}
    \centering
    \includegraphics[width=\textwidth]{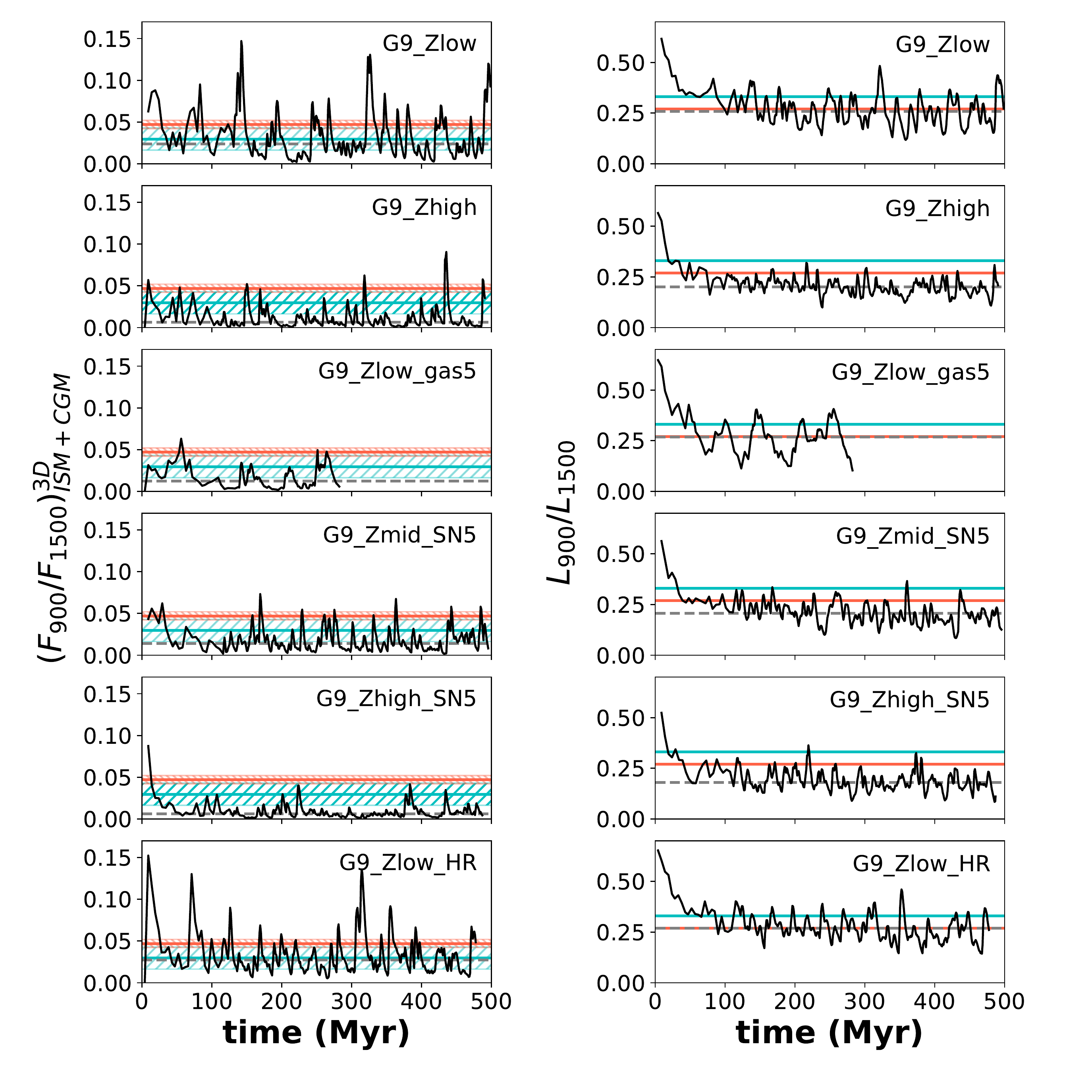}
    \caption{The intrinsic luminosity ratio ($L_{900}/L_{1500}$) and the IGM-corrected flux ratio ($\left( F_{900} / F_{1500} \right)_{\rm ISM+CGM}^{\rm 3D}$), based on the wavelength definition of \citetalias{marchi17}. The grey dashed lines indicate the median values in each simulation. The observational estimates of \citetalias{marchi17} and \citetalias{steidel18} are shown as the cyan and red lines, respectively. The flux ratio in the fiducial run ($\langle\left(F_{900} / F_{1500} \right)_{\rm ISM+CGM}^{\rm 3D}\rangle=0.023$) is similar to the results derived from the galaxies with $M_{\rm UV}\simeq-20$ in \citetalias{marchi17}.  $L_{900}/L_{1500}$ from the metal-poor runs predicts  $L_{900}/L_{1500}=0.27$, which is reasonably consistent with the results of  \citetalias{marchi17} and \citetalias{steidel18}. The runs with lower metallicity tend to have a higher $\left( F_{900} / F_{1500} \right)_{\rm ISM+CGM}^{\rm 3D}$. }
    \label{fig:flux_ratio}
\end{figure*}

In observations, it is very difficult to derive absolute escape fractions because the properties of dust at high redshift are not well constrained. To circumvent the uncertainty, the relative escape fraction is used instead, which can be given as \citep[e.g.,][]{siana07,vanzella10,marchi17} 
 \begin{align}
    \label{eq:fescrel}
         f_{\text{esc,rel}} & \equiv \frac{F_{900}/F_{1500}}{L_{900}/L_{1500}}\times\exp\left(\tau_{\rm HI}^{\rm IGM} \right)  
\end{align}
  where $F_{\lambda}$ is the observed flux measured at some wavelength $\lambda$,  $L_{\lambda}$ is the intrinsic flux, and $\tau_{\rm HI}^{\rm IGM}$ is the optical depth to the LyC photon due to the IGM at 900 \r{A}. Here, the last term in Eq.~\ref{eq:fescrel} corrects for the absorption of ionizing radiation due to  hydrogen in the IGM \citep[e.g.,][]{inoue14}.  Alternatively, the relative escape fraction may be computed as follows \citep[e.g.,][]{steidel18}
   \begin{equation}
     f_{\rm esc,rel}' \equiv \frac{F_{900}/F_{1500}}{L_{900}/L_{1500}}\exp \left( \tau_{\rm HI}^{\rm IGM}\right) \times10^{0.4\left[A(900)-A(1500)\right]},
 \end{equation}
 which additionally considers the effect of differential dust attenuation ($A[900]-A[1500]$).

\begin{table*}
    \centering
    \begin{tabular}{c|c|c|c|c|c|c}
         \hline
         Simulation & $(F_{900}/F_{1500})_{\rm ISM+CGM}^{\rm 3D}$ & $L_{900}/L_{1500}$ & \fesctdL & $\langle f_{\rm esc,rel}^{\rm 3D} \rangle_{\mathcal{L}}$ (M17) & $\langle f_{900}^{\rm 3D} \rangle_{\mathcal{L}}$ & $\langle f_{1500}^{\rm 3D} \rangle_{\mathcal{L}}$  \\
         \hline
         G9\_Zlow & $0.024_{-0.011}^{+0.020}$  & $0.258_{-0.048}^{+0.057}$   & 
         $0.104$ &
         $0.143$ & 0.088 & 0.573\\
           G9\_Zhigh  & $0.006_{-0.003}^{+0.007}$ & $0.200_{-0.022}^{+0.030}$         & 
           $0.012$ &
           $0.055$  & 0.009 & 0.128    \\
        
         G9\_Zlow\_gas5 & $0.013_{-0.007}^{+0.013}$ & $0.269_{-0.068}^{+0.071}$ &
         $0.048$ &
         $0.070$ &0.029& 0.414  \\
          G9\_Zmid\_SN5 & $0.014_{-0.006}^{+0.009}$ &
          
          $0.207_{-0.031}^{+0.041}$ &
          $0.046$ &
          $0.086$ &0.037 & 0.408 \\ 
       
         G9\_Zhigh\_SN5 & $0.006_{-0.002}^{+0.004}$ & $0.182_{-0.023}^{+0.039}$    &
         $0.015$ &
         $0.045$  & 0.010 & 0.210         \\
          G9\_Zlow\_HR & $0.027_{-0.010}^{+0.014}$ & $0.269_{-0.041}^{+0.047}$               &
         $0.100$ &
         $0.126$  & 0.074 & 0.567   \\
         \hline
    \end{tabular}
    \caption{The median values of the three-dimensional IGM-corrected flux ratio ($(F_{900}/F_{1500})_{\rm ISM+CGM}^{\rm 3D}$),  intrinsic luminosity ratio ($L_{900}/L_{1500}$),  luminosity-weighted absolute escape fraction (\fesctdL),  luminosity-weighted relative escape fraction ($\langle f_{\rm esc,rel}^{\rm 3D} \rangle_{\mathcal{L}}$) computed based on Eq.~\ref{eq:fescrel2}, luminosity-weighted escape fraction of photons with wavelength $\sim 900\,\AA$ ($f_{900}\equiv F_{900}/L_{900}$), and escape fraction of photons with wavelength $\sim 1500\,\AA$ ($f_{1500}\equiv F_{1500}/L_{1500}$). The 25\% and 75\% percentiles of each quantity are also presented. All quantities are measured at $\tsim>150\,{\rm Myr}$. The unit of the intrinsic flux and IGM-corrected flux used is ${\rm erg}\,{\rm s^{-1}}\,{\rm Hz^{-1}}$.}
    \label{tab:fescrel}
\end{table*}

 To avoid any possible confusion due to the different definitions, we directly compare the ratio of the fluxes at 900 \r{A} and 1500 \r{A}. We note that the observed flux is attenuated not only by the gas in the dark matter halo (i.e. ISM plus CGM) but also by the IGM \citep{steidel01,siana07}: 
 \begin{align}
         \frac{F_{900}}{F_{1500}} =  \frac{L_{900}}{L_{1500}} & \times 10^{-0.4\left[ A(900)-A(1500)\right]} \nonumber\\ 
         & \times \exp(-\tau_{\rm HI}^{\rm IGM})\times \exp(-\tau_{\rm HI}^{\rm ISM+CGM}).
 \end{align}
Because we do not model the IGM in our simulations, we use the IGM-corrected flux ratio, 
\begin{equation}
\left( \frac{F_{900}}{F_{1500}} \right)_{\rm ISM+CGM} \equiv   \left( \frac{F_{900}}{F_{1500}} \right) \exp \left( \tau_{\rm HI}^{\rm IGM}\right),
\end{equation}
which is identical to the definition of $\left( f_{900} / f_{1500}  \right)_{\rm out}$ in \citet{steidel18}. In simulations, calculating the flux ratio is straightforward by attenuating the intrinsic spectrum from each star particle with gas and dust in the dark matter halo in a similar way to Eqs.~\ref{eq:fesc_def1}--\ref{eq:fesc_def2}.
 
Fig.~\ref{fig:flux_ratio} shows that the median flux ratio at the virial radius, $\left( F_{900} / F_{1500} \right)_{\rm ISM+CGM}^{\rm 3D}$, is  $ 0.024_{-0.011}^{+0.020}$ in the fiducial run, where the error indicates the interquartile range. In contrast, the flux ratio in the \texttt{G9\_Zhigh} run is $ 0.006_{-0.003}^{ +0.007}$, indicating that the flux ratio decreases with increasing metallicity.  This is partly because the intrinsic flux ratio, $L_{900}/L_{1500}$, is lower in more metal-rich stellar populations (0.258 vs. 0.200, see Table~\ref{tab:fescrel}). More importantly, as studied in Section 3.3, the attenuation due to neutral hydrogen in more metal-rich galaxies is stronger because the young stars are trapped in star-forming clumps for a longer time. Interestingly, if the results are compared between \texttt{G9\_Zhigh} and  \texttt{G9\_Zhigh\_SN5}, the flux ratio is lower in the \texttt{G9\_Zhigh\_SN5} run where the enshrouded timescale is actually shorter because of the strong SN feedback. This happens because feedback prevents gas from turning into stars and increases the column density of neutral hydrogen in the disc, as shown in  Fig.~\ref{fig:fesc_distance}. For the same reason, gas-rich galaxies with low metallicity (\texttt{G9\_Zlow\_gas5}) show a $\left( F_{900} / F_{1500} \right)_{\rm ISM+CGM}^{\rm 3D}= 0.013_{-0.007}^{+0.013}$, which is lower than that of the fiducial run.

Recently, \citet[][hereafter M17]{marchi17} combined 33 galaxies with $M_{\rm UV}\approx-20$ at $z\sim4$ and obtain a observed flux ratio of $F_{900}/F_{1500}=0.008\pm0.004$, where $F_{900}$ and $F_{1500}$ are measured at [880\r{A}, 910\r{A}]  and [1420\r{A}, 1520\r{A}], respectively. Assuming the mean IGM transmission of $\left< \exp \left(-\tau_{\rm HI}^{\rm IGM} \right)\right>=0.27$, \citetalias{marchi17}  found $\left(F_{900}/F_{1500}\right)_{\rm ISM+CGM}=0.030$. \citet[][hereafter S18]{steidel18} used 124 faint ($M_{\rm UV}\sim -19$) galaxies at $z\sim3$, and obtained a higher value of $F_{900}/F_{1500}=0.021\pm0.002$, where $F_{1500}$ is measured at slightly different wavelength ranges [1475\r{A}, 1525\r{A}]. Adopting the IGM transmission of $\left< \exp \left(-\tau_{\rm HI}^{\rm IGM} \right)\right>=0.443$, which is appropriate for $z\sim3$, \citetalias{steidel18} concluded that the mean IGM-corrected flux ratio is 0.047, suggesting that the ratio depends on the luminosity of the galaxy sample. 

We note that the flux ratios obtained from our simulations with low to intermediate metallicity (\texttt{G9\_Zlow}, \texttt{G9\_Zlow\_gas5}, and \texttt{G9\_Zmid\_SN5}) are consistent with the observational estimates of \citetalias{marchi17} within the errors. However, compared with the estimates of \citetalias{steidel18}, our flux ratios are lower, which is likely due to the fact that the \citetalias{steidel18} sample represents more metal-poor systems ($Z\sim0.001$). Our metal-rich runs predict the flux ratios that are lower than \citetalias{marchi17} and \citetalias{steidel18} but are more consistent with the results obtained from the more luminous sample of \citet{grazian16} ($M_{\rm UV}\sim-21$,  $\fescrel\la2\%$)\footnote{If we take the ratio of stacked fluxes of 37 galaxies at $U$ and $R$ bands, $F_{\rm R}/F_{\rm U}=545.1$, as $F_{1500}/F_{900}$ and if we use the mean IGM transmission from the literature ($\langle \exp{(-\tau_{900}^{\rm IGM})} \rangle=0.28$), their flux ratio is  $\left(F_{900}/F_{1500}\right)_{\rm ISM+CGM}=0.007$.}. In light of these comparison, we argue that the low flux ratio derived in bright galaxies is primarily due to higher metallicities.

In Table~\ref{tab:fescrel}, we compare the relative and absolute escape fractions. The relative escape fraction is computed adopting the definition of wavelength given by \citetalias{marchi17} as follows:
 \begin{align}
    \label{eq:fescrel2}
         f_{\text{esc,rel}}^{\rm 3D} & \equiv \frac{F_{900, \rm vir}/F_{1500, \rm vir}}{L_{900}/L_{1500}},
\end{align}
where $F_{\rm vir}$ is the attenuated flux measured at the virial radius. We find that the relative escape fractions are quite similar to \fesctd\ in the low metallicity runs, whereas they diverge at higher metallicities. Because the relative escape fraction is essentially the ratio of the escape fractions at two different wavelengths, $f_{900}/f_{1500}$, the fact that the relative escape fraction is significantly higher than \fesctd\ indicates that the UV photons with $\lambda\approx 1500\, \AA$ are more efficiently absorbed  by dust in the metal-rich runs. Indeed, $\sim60\%$ of the UV photons escape from the metal-poor runs, whereas only  $\sim10$--$20\,\%$ of them manage to leave their  dark matter halos in the metal-rich cases (see $\langle f_{1500}^{\rm 3D} \rangle_{\mathcal{L}}$ in Table~\ref{tab:fescrel}). The relative escape fraction in the metal-poor galaxies is $\fescreltd \approx 14\,\%$,  which is in between the results of \citet{marchi17} ($\fescrel\approx8$--$9\,\%$) and \citet{steidel18} ($\fescrel\sim20\,\%$\footnote{ We compute the average relative escape fraction, defined as in Equation~\ref{eq:fescrel}, by comparing the observed and intrinsic flux ratio ($\langle f_{900}/f_{1500}\rangle_{\rm obs}=0.021$, $\langle f_{900}/f_{1500}\rangle_{\rm int}=0.28$) and by adopting the IGM+CGM transmission of 0.368.}).

One may wonder at this point whether comparing the properties of our simulated galaxies to those of the LBGs is appropriate, given that their host dark matter halo mass differs by an order of magnitude. Indeed, if the gas mass is increased to the level of LBGs, it would lower the escape fraction (Section 3.4), and thus there is a possibility that our simulations are under-estimating $f_{\text{esc,rel}}^{\rm 3D}$ by a few percents (e.g., compare the \citealt{marchi17} results with \texttt{G9\_Zlow\_gas5}). However, we note that the dependency of $f_{\text{esc,rel}}^{\rm 3D}$ on metallicity should still be valid and that the high metallicity is needed to explain low escape fractions of the luminous galaxies by \citet{grazian16}. Another important difference between the two samples is the star formation rate, but as we will show in Section 4.3, we find little correlation between the star formation surface density and the escape fraction, indicating that the difference in star formation rates is unlikely to make a significant impact on our conclusions.

 Finally, it is worth emphasising that the luminosity-weighted absolute escape fraction measured at [880\,\AA, 910\,\AA], $\left< f_{900}^{\rm 3D}\right>_{\mathcal L}$, is systematically lower by $\sim  15\,\%$ than \fesctdL\ because the absorption cross-section due to neutral hydrogen is the largest near the Lyman edge \citep[see also][]{kimm19}. For example, the luminosity-weighted $f_{900}$ in the fiducial run is  8.8\,\%, whereas $\left<\fesctd\right>_{\mathcal L}$ between 150 to 500 Myr is 10.4\,\%. Likewise, the run that has the minimum $\left<\fesctd\right>_{\mathcal L}$ of  1.2\,\% (\texttt{G9\_Zhigh}) shows $f_{900}= 0.9\,\%$. Therefore, the observationally derived escape fractions should be carefully compared with the theoretical value required to reionize the Universe at $z\sim 6$.

\subsection{Comparison with other numerical studies}

Previous studies that measured the theoretical escape fractions have mainly focused on galaxies at $z\ga 6$ to determine their contribution to reionization of the Universe \citep[e.g.][]{wise14,paardekooper15,ma16,xu16,trebitsch17,rosdahl18}. However, few attempts have been made to examine the escape fraction of massive galaxies embedded in $M_h\simeq10^{11-12}M_\odot$ at $z\sim3$, which are the main target in this study.  A few exceptions include the studies by \citet{gnedin08,yajima11,kim13b}. Using cosmological radiation-hydrodynamics simulations with the maximum resolution of 50 pc at $z=3$, \citet{gnedin08} showed a positive correlation between the escape fraction and the halo mass. By contrast, based on the post-processing of hydrodynamic simulations, \citet{yajima11} argued that the escape fraction is lower in more massive galaxies, although both studies suggest low $\fesctd$ of $\la10\,\%$ for the galaxies with $M_h\simeq10^{11-12}M_\odot$ at $z\sim3$, similar to our findings.

\cite{kim13b} simulated the propagation of ionizing radiation in an isolated disc galaxy with a halo mass of $M_h=2.3\times10^{11}M_\odot$ adopting the maximum resolution of 3.8 pc. They showed that the escape fraction varies from $0.8\%$ to $5.9\%$ over time, with the temporal average of $\left< \fesctd \right>=1.1\,\%$. Given that initially the gas of the galaxy is more metal-poor ($0.003\,Z_\odot$) than our fiducial model, their predicted escape fractions are significantly lower than our expectations. Because numerical methods as well as initial conditions are dissimilar, it is difficult to make a direct comparison, but we note that there are several important differences that could result in higher escape fractions in this work. First, our simulations include forms of strong feedback, e.g. Ly$\alpha$ pressure and (boosted) mechanical feedback, which allows for more LyC photons to escape. Second, the stellar SEDs in \citet{kim13b} assumed single stellar evolution which likely led to a lower $f_{\text{esc}}$ than in our results, for which we used binary SEDs \citep[e.g.,][]{ma16, rosdahl18}.

It is also interesting to compare our results with those from recent cosmological radiation-hydrodynamics simulations. Using strong SN feedback and runaway stars, \citet{kimm14} showed that there is a weak negative correlation between halo mass and escape fraction. Their predicted escape fractions in intermediate-mass halos with $10^{10} < M_{\rm h} < 10^{11}\,\msun$ are slightly higher than $\fesctd=10\%$, which is likely due to the different star formation models used. In \citet{kimm14}, stars form once the density of a converging flow is greater than $n_{\rm H} = 100\, {\rm cm^{-3}}$ with a fixed $\epsilon_{\rm ff}=0.02$, whereas in this work they form preferentially in locally gravitationally well bound, dense environments ($n_{\rm H} \sim 10^4\, {\rm cm^{-3}}$). Thus it is more difficult for young stars in our simulations to disrupt their birth clouds, leading to lower escape fractions. In contrast, based on the Renaissance simulations, \citet{xu16} concluded that although some galaxies in massive halos with $ M_h \sim 10^{9.25}\,\msun$ are efficient LyC leakers ($\fesctd\sim15\%$), the escape fractions in halos of mass $\sim 10^{9}\,\msun$ are generally low ($\fesctd\la 5\,\%$), which is perhaps due to the absence of strong stellar feedback such as Ly$\alpha$ pressure. Adopting  mechanical SN feedback, the same model as here but without the extra boost, \citet{trebitsch18} found that galaxies with a halo mass of $M_h\simeq5\times 10^9M_\odot$ show $\fesctd\approx 6-8\,\%$, depending on the presence of feedback from black holes. The simulated galaxies with strong feedback in the SPHINX simulations \citep{rosdahl18} also show \fesctd\ of $7$--$10\%$ in halos of mass $10^8 \la M_h / M_\odot\la10^{10}$ at $z=6$,  which is similar or slightly lower than those from our relatively metal-poor galaxies.

\subsection{Correlation with star formation surface density and outflow rates}

\begin{figure}
    \centering
    \includegraphics[width=8.3cm]{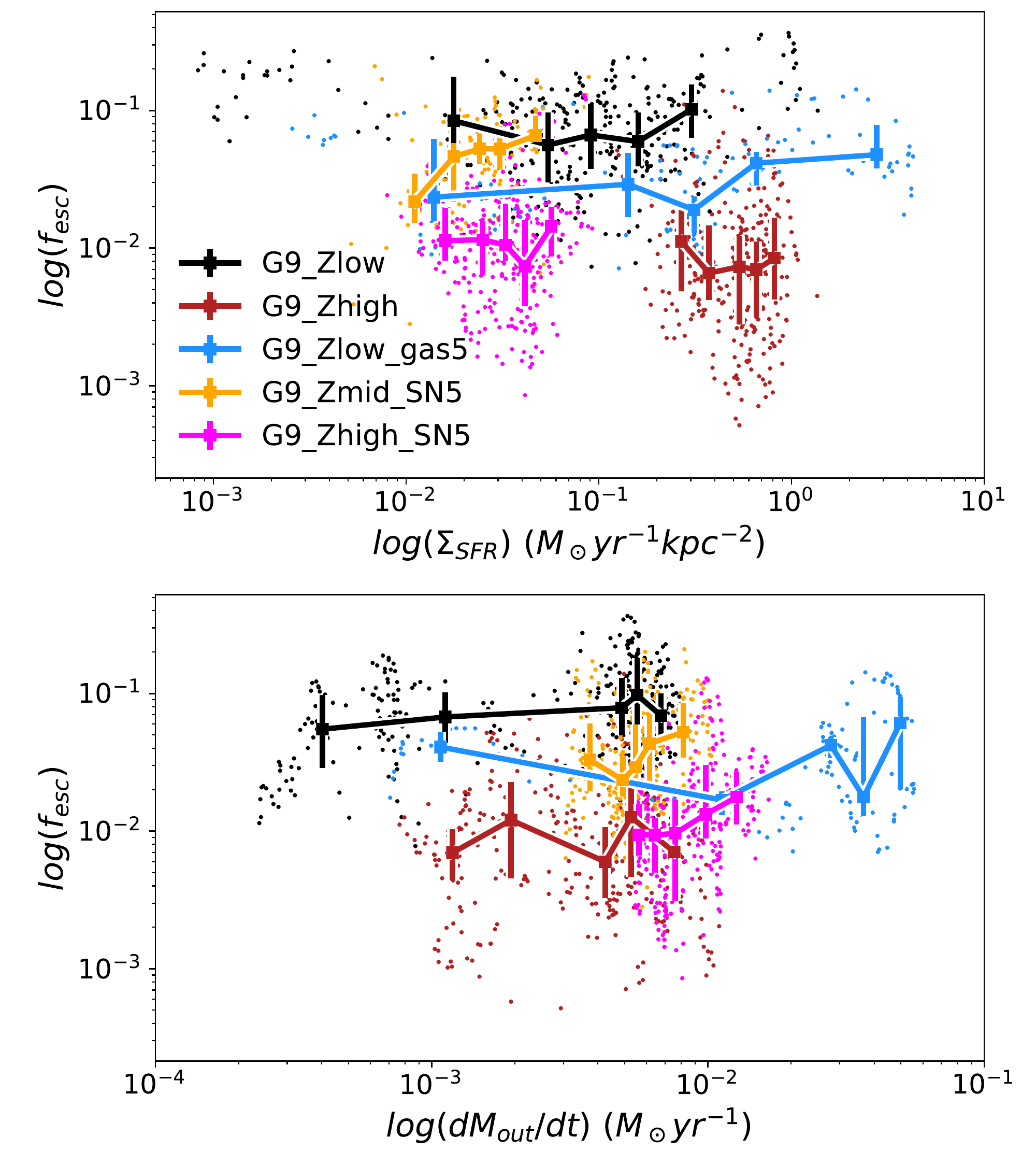}
    \caption{Correlation between average escape fractions and star formation surface density ($\Sigma_{\rm SFR}$; upper panel) and outflow rates ($dm_{\rm out}/dt$; lower panel). $\Sigma_{\rm SFR}$ is measured from the total star formation rate averaged over 10 Myr within the half-mass radius of the galaxies at $ \tsim > 150 \, {\rm Myr}$ and the outflow is measured at $|z|=10\,{\rm kpc} $ from disc plane. Again, different colour-codes correspond to different simulations, as indicated in the legend. The error bars indicate the interquartile range of the distributions. The correlation between the two quantities in both panels appears to be weak even when the data from the run with weak feedback, i.e. \texttt{G9\_Zhigh}, is excluded.}
    \label{fig:sfrd_fesc}
\end{figure}

To constrain the escape of LyC photons from models of the reionization history of the Universe, \citet{sharma17} and \citet{naidu19} conjectured that \fesctd\ is correlated with the star formation rate density. This is motivated by the idea that strong outflows would develop as a result of vigorous star formation activities, carving out low-density channels through which the LyC photons easily escape. Indeed, some LyC leakers appear to be compact and actively star-forming \citep[e.g.,][]{vanzella18,izotov18}, supporting the idea that the escape may be related to star formation rate density \citep{naidu19}. To examine this hypothesis, we plot the relation between the luminosity-weighted escape fraction and star formation surface density ($\Sigma_{\rm SFR}$) from each snapshot in Fig.~\ref{fig:sfrd_fesc}. Here, $\Sigma_{\rm SFR}$ is computed as the total star formation rate  averaged over 10 Myr within the stellar half-mass radius ($r_{\rm eff,m}$) divided by $\pi r_{\rm eff,m}^2$.

Fig.~\ref{fig:sfrd_fesc} shows that there is no clear correlation between \fesctd\ and $\Sigma_{\rm SFR}$ in our simulations. In the \texttt{G9\_Zlow} run, the escape fractions are clustered around $\sim 10\,\%$, even though the surface density varies from $0.001$ to $1 \,M_\odot\,{\rm yr}^{-1}\, {\rm kpc}^{-2}$. Similarly, metal-rich galaxies with strong SN feedback show a flat distribution of \fesctd\ as a function of $\Sigma_{\rm SFR}$. This is because the fluctuating behavior of the SFR and \fesctd\ of a galaxy is asynchronous (Fig.~\ref{fig:fig3}). In particular, we emphasise that \fesctd\ fluctuates within short time intervals comparable to the time delay ($\sim5$--$ 10\,{\rm Myr}$) between the SFR and \fesctd\ of a galaxy, which results in little correlation. Even when we combine the results from different metallicities, except for \texttt{G9\_Zhigh}, which uses stellar feedback parameters that fail to reproduce luminosity functions at high redshift \citep{rosdahl18}, the correlation still seems very weak, in conflict with the assumption used in \citet{sharma17} or \citet{naidu19}. Galaxies can have high \fesctd\ even when $\Sigma_{\rm SFR}$ is low ($\la 0.1 \,M_\odot\,{\rm yr}^{-1}\, {\rm kpc}^{-2}$) or low \fesctd when $\Sigma_{\rm SFR}$ is high ($\ga 0.1 \,M_\odot\,{\rm yr}^{-1}\, {\rm kpc}^{-2}$).

We also examine the relation between the escape fraction and outflow rates in the bottom panel of Fig.~\ref{fig:sfrd_fesc}. We measure the outflow rates at 10 kpc (nearly 0.1$R_{\rm vir}$) above or below the disc mid-plane. Again, we find little correlation between the two properties because the outflow rates  vary smoothly (Fig.~\ref{fig:vert}), unlike escape fractions. While the escape fractions are sensitive to very young stars with $\tage\la 5\,{\rm Myr}$, outflows can be launched by SNe until $\approx 8\,\msun$ stars evolve off the main sequence at $\tage\approx 40\, {\rm Myr}$. Moreover, early SNe going off at $\tage \la 10\,{\rm Myr}$ do not necessarily drive stronger outflows than the ones exploding at later times because the former are likely to explode in denser environments where the radiative cooling is expected to be efficient. We find that the correlation is unclear even for the outflows measured at different regions ($|z|=2\,{\rm kpc}$ or $20\,{\rm kpc}$). These results suggest that instantaneous outflow rates alone may not be best to select potential LyC leakers \citep[c.f.,][]{heckman11,chisholm17}.

\subsection{Resolution convergence}

\begin{figure}
    \centering
    \includegraphics[width=8.3cm]{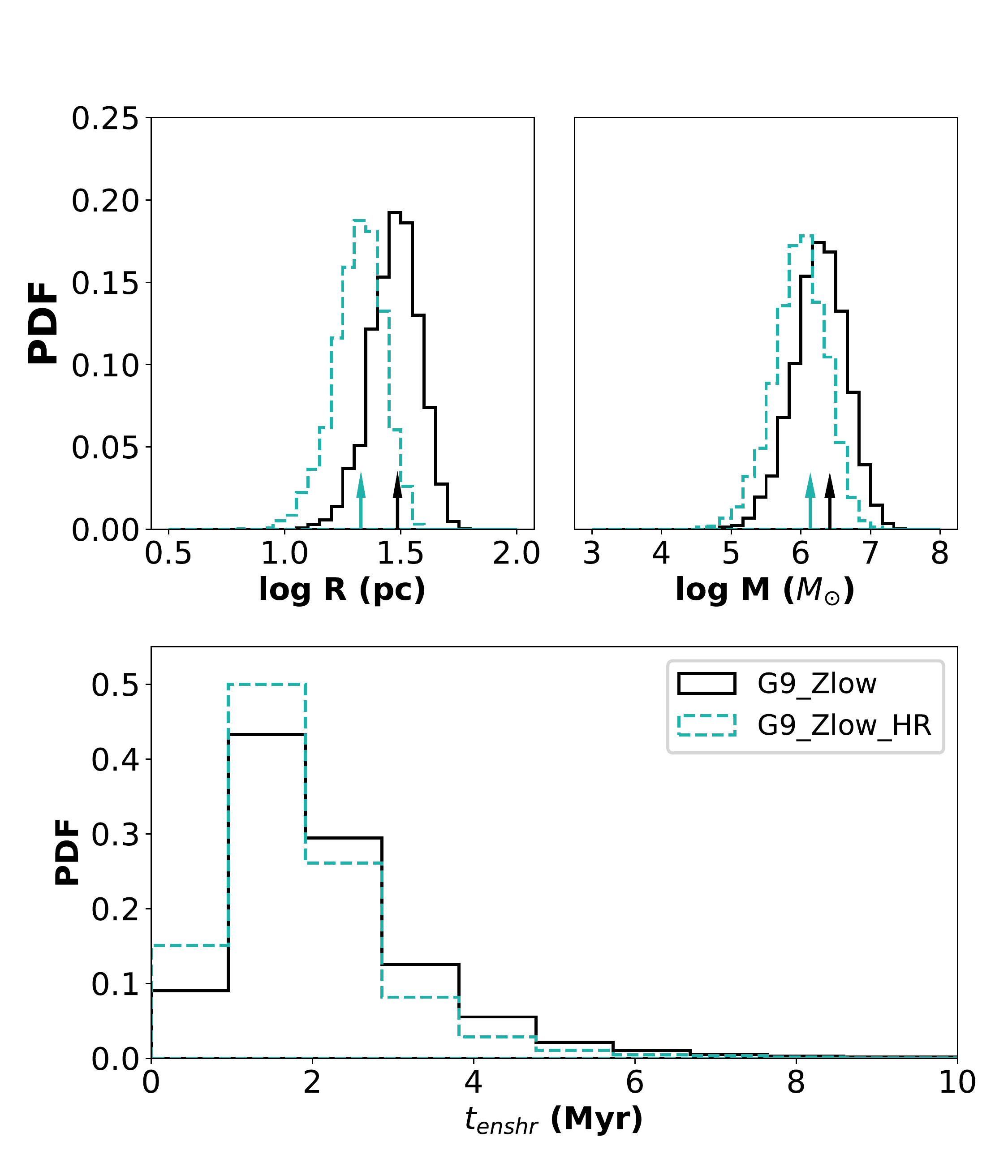}
    \caption{Properties of gas clumps in two different resolution runs, \texttt{G9\_Zlow} (9.2 pc) and \texttt{G9\_Zlow\_HR} (4.6 pc). The gas clumps in the higher resolution simulation are smaller and less massive. As a result, the clumps are more easily disturbed by stellar feedback, leading to a shorter enshrouded timescale.}
    \label{fig:add1}
\end{figure}

To test how well our results are converged with resolution, we perform an additional simulation with a higher maximum resolution of $\Delta x=4.6\,{\rm pc}$ for the fiducial case. 

Fig.~\ref{fig:add1} shows that gas clumps become somewhat less massive ($1.2\times10^6M_\odot$) and smaller in radius (20 pc) in the \texttt{G9\_Zlow\_HR} run than in the fiducial run with 9 pc resolution ($2\times10^6M_\odot$ and  30 pc). The gas in the \texttt{G9\_Zlow\_HR} disc is fragmented more efficiently, producing a larger number of star-forming clumps. We find that the enshrouded timescale becomes shorter from $\sim2.3$ Myr to $\sim1.9$ Myr, on average, indicating that the clumps get more easily disrupted because of strong feedback. The escape fraction measured on clump scales ($\sim$40 pc) is also slightly increased to 0.70 in the higher resolution run from 0.64 in the fiducial case. However, the absorption of LyC photons on $\sim 0.1$--$1$ kpc scales in \texttt{G9\_Zlow\_HR} turns out to be slightly more enhanced, compensating the differences on galactic scales. This can be attributed to the fact that the smaller clouds get disrupted early and that SNe redistribute the gas to the ISM.  As a result, not only the star formation rate but also the luminosity-weighted escape fraction in the \texttt{G9\_Zlow\_HR} run are found to be nearly the same as those in the \texttt{G9\_Zlow} run (see Fig.~\ref{fig:fig3} and Table~\ref{tab:fesc}). 
A similar trend is also found in \citet[][Fig. C1]{kimm14}, where the results are reasonably converged at a 4 pc resolution. Given that the difference in the escape fraction estimated from different resolution runs is not very significant, it is unlikely to change our main conclusions regarding the dependence of the escape fractions on metallicity and gas fraction, but this should be tested with even higher resolution simulations in the near future.

\section{Summary and conclusions}
 
To study the origin of the inefficient leakage of LyC photons from  massive star-forming galaxies at $z\sim3$, we investigate the propagation of LyC photons from stellar populations in isolated disc galaxies embedded in a $10^{11}\,\msun$ dark matter halo. For this purpose, we employed strong stellar feedback, in the form of mechanical SN explosions and Ly$\alpha$ pressure, which can self-regulate star formation in the galaxies. Our findings are summarised as follows:

\begin{itemize}
    \item [1.]  We find that the luminosity-weighted average escape fraction of LyC photons ($\left<\fesctd\right>_\mathcal{L}$) in our fiducial run with low metallicity ($Z=0.002$) is 10.4\%, but it decreases significantly with increasing gas metallicity. Only $\left<\fesctd\right>_\mathcal{L}\approx 1\,\%$ of LyC photons escape from our metal-rich galaxies ($Z=0.02$). In contrast, when the mass of the gas disc is increased by a factor of 5 (motivated by the upper limit of the gas fraction inferred from high-$z$ observations), the escape fraction is mildly decreased to $\left<\fesctd\right>_\mathcal{L} = 4.8\,\%$. Our results thus suggest that the low escape fraction measured from the massive galaxies at high redshift, compared to what reionization models typically assume ($\sim 10$--$20\,\%$),  is likely due to higher metallicities. 
    
    \item[2.]  In metal-poor galaxies, strong radiation feedback efficiently disrupts the star-forming clouds, and the majority of young stars are no longer enshrouded by their birth clouds within $\tensh\sim 2\,{\rm Myr}$. We measure that roughly a half of the LyC photons are absorbed on  local scales ($50-100\, {\rm pc}$), and the other half is absorbed by the ISM ($\la 2\,{\rm kpc}$).
    
    \item[3.] The LyC photons from the metal-rich galaxies are absorbed by the clumps for a longer time  ($ \tensh\sim 10\,{\rm Myr}$, due to weaker radiation field, enhanced metal cooling, and more effective destruction of Ly$\alpha$ photons. The longer enshrouded timescale in the metal-rich system leads to a lower galactic escape fraction than in the fiducial run. In addition, as the intrinsic ionizing emissivity from metal-rich stars falls more rapidly than that of the metal-poor stars, the escape fraction in the metal-rich galaxies becomes significantly reduced compared to the $Z=0.002$ case.  
    
    \item[4.]  Increasing the gas mass by a factor of five has little impact on the enshrouded timescale of young stars, although star formation efficiencies and clump masses ($\epsilon_{\rm clump}= 1.2 \,\%$, $M_{\rm cloud}= 3.5\times10^6M_\odot$) in the gas-rich disc are slightly increased compared with those in the fiducial run ($\epsilon_{\rm clump}=0.5\,\%$, $M_{\rm cloud}= 1.9\times10^6M_\odot$). While the gas-rich disc shows that a similar fraction of ionizing photons are absorbed at the clump scale ($\sim40 \,{\rm pc}$), a larger fraction is absorbed by the gaseous disc ($d_{\rm star} \ga 1\,{\rm kpc}$) which is more extended because of vigorous star formation activities and associated outflows, resulting in an overall lower escape fraction in the gas-rich disc.
    
    \item [5.] We find that the luminosity-weighted average escape fractions from the metal-rich runs are very similar ($\fesctdL\approx1\%$), regardless of whether the frequency of SN explosions is boosted by a factor of five or not. Even though young stars in the run with boosted SN feedback escape from the dense birth clouds earlier ($\tensh\sim  5\,{\rm Myr}$) than without, powerful feedback thickens the disc, increasing the column density of neutral hydrogen at $|z|\ga 1 \, {\rm kpc}$. As a result, the escape fractions are rather insensitive to the strength of SN feedback for the metal-rich, massive disc galaxies examined in this study. 
    
    \item[6.] Our simulated galaxies with metallicity of $Z=0.002$--$0.006$ show a similar flux ratio $\left(F_{900}/F_{1500}\right)_{\rm CGM+IGM}^{\rm 3D}\sim 0.01$-- $0.03$ as the observations of $M_{\rm UV}\sim-20$ galaxies \citep{marchi17}, but it is lower than the fainter ($M_{\rm UV}\sim-19$) and more metal-poor ($Z=0.001$) sample by \citet{steidel18}. In contrast, the low escape fraction estimated from UV bright galaxies with $M_{\rm UV}\sim-21$ \citep[$\fescrel\la2\%$,][]{grazian16} is similar to those of our metal-rich galaxies, supporting the claim that the low escape fraction in massive and bright systems is mainly due to metal enrichment.  
    
    \item [7.]  We find that the star formation surface density does not correlate well with the escape fraction. This is because the escape of LyC photons typically peaks $\sim 5$--$20\,{\rm Myr}$ after the peak in star formation, which is comparable to the fluctuation timescale of the escape fraction. The escape fractions are also uncorrelated with the galactic outflow rates because they vary smoothly as SNe explode over the timescale of $\sim 40\, {\rm Myr}$, whereas the escape fraction fluctuates on much shorter timescales ($t<10\, {\rm Myr}$).
    
\end{itemize}

We show that the escape fractions are sensitive to the gas metallicity of massive galaxies at high redshift. Massive halos are more metal-rich than low-mass halos, which naturally suggests that the escape fractions are negatively correlated with dark matter halo mass. Admittedly, however, our results are based on a single galaxy in an isolated environment, and cosmological zoom-in simulations that specifically target the evolution of massive galaxies ($M_h\ga10^{11}\,M_\odot$) will be required to draw a statistically meaningful conclusion. At the same time, future observational efforts to measure LyC flux need to be extended to fainter galaxies to test our hypothesis on the relationship between the metallicity and escape fraction.

\section*{Acknowledgements}

 We thank the referee, Nick Gnedin, for constructive comments. We are grateful to Jeremy Blaizot, Maxime Tresbitsch, Julien Devriendt, Adrianne Slyz, and Sandro Tacchella for useful discussion, and Kearn Grisdale for sharing the parameters for \texttt{PHEW}. TK was supported in part by the Yonsei University Future-leading Research Initiative (RMS2-2019-22-0216) and in part by the National Research Foundation of Korea (NRF-2017R1A5A1070354 and  NRF-2020R1C1C100707911). This work was supported by the Supercomputing Center/Korea Institute of Science and Technology Information with supercomputing resources including technical support (KSC-2018-CRE-0099). The results of this research have been achieved using the PRACE Research Infrastructure resource JUWELS based in J\"{u}lich, Germany (project 2018184362). We are grateful for the excellent technical support provided by the JUWELS staff. JR acknowledges support from the ORAGE project from the Agence Nationale de la Recherche under grand ANR-14-CE33-0016-03. 
 This work was also performed using the DiRAC Data Intensive service at Leicester, operated by the University of Leicester IT Services, which forms part of the STFC DiRAC HPC Facility (www.dirac.ac.uk). The equipment was funded by BEIS capital funding via STFC capital grants ST/K000373/1 and ST/R002363/1 and STFC DiRAC Operations grant ST/R001014/1. DiRAC is part of the National e-Infrastructure.

 \section*{Data availability}
 The data underlying this article will be shared on reasonable request to the corresponding author.




\bibliographystyle{mnras}
\bibliography{refs} 


\bsp	
\label{lastpage}
\end{document}